\documentclass[a4paper,12pt]{article}

\usepackage[margin=3cm,footskip=1cm]{geometry}

\usepackage[T1]{fontenc}

\usepackage{amsmath,amssymb,amsthm,bm,mathtools,xspace,booktabs,url}
\usepackage{graphicx,etoolbox}

\usepackage[shortlabels]{enumitem}
\setlist{
	listparindent=\parindent,
	parsep=0pt,
}

\usepackage[round]{natbib}

\AtBeginEnvironment{thebibliography}{%
	\small
	\setlength\itemsep{0.75em plus 0.5em minus 0.75em}%
}

\usepackage{caption}
\usepackage[raggedright]{titlesec}

\numberwithin{equation}{section}

\usepackage{authblk}

\makeatletter
\newcommand\CorrespondingAuthor[1]{%
	\begingroup%
	\def\@makefnmark{}%
	\footnotetext{Corresponding author: #1}%
	\endgroup%
}
\makeatother

\makeatletter
\renewenvironment{abstract}{%
	\small%
	\providecommand\keywords{%
		\par\medskip\noindent\textit{Keywords:}\xspace}%
	\begin{center}%
		\bfseries \abstractname\vspace{-.5em}\vspace{\z@}%
	\end{center}%
	\quote%
}{\endquote}
\makeatother

\usepackage[above,below,section]{placeins}

\usepackage[load-configurations=abbreviations]{siunitx}
\usepackage{lipsum}

\newtheoremstyle{example}
{\topsep} {\topsep}%
{\upshape}
{}
{\itshape}
{.}
{1em}
{\thmname{#1}\thmnumber{ #2 }\thmnote{#3}}

\theoremstyle{example}

\newcommand{\di}{\mathrm{d}}
\newcommand{\E}{\mathrm{E} \,}

\usepackage{chngcntr}

\begin{document}

\title{Modelling spine locations on dendrite trees using inhomogeneous Cox  point processes}

\author{Heidi S. Christensen}
\author{Jesper M{\o}ller} 
\affil{Department of Mathematical Sciences, Aalborg University}	
\date{}

\maketitle

\begin{abstract}
Dendritic spines, which are small protrusions on the dendrites of a neuron, are of interest in neuroscience as they are related to cognitive processes such as learning and memory. 
We analyse the distribution of spine locations on six different dendrite trees from mouse neurons using point process theory for linear networks. Besides some possible small-scale repulsion, { we find that two of the spine point pattern data sets may be described by inhomogeneous Poisson process models}, while the other point pattern data sets exhibit clustering between spines at a larger scale. To model this we propose an inhomogeneous Cox process model constructed by thinning a Poisson process on a linear network with retention probabilities determined by a spatially correlated random field. 
For model checking we consider network analogues of the empirical $F$-, $G$-, and $J$-functions originally introduced for inhomogeneous point processes on a Euclidean space. 
The fitted Cox process models seem to catch the clustering of spine locations between spines, but also posses a large variance in the number of points for some of the data sets causing large confidence regions for the empirical $F$- and $G$-functions.

\keywords Empirical summary functions; linear networks; random fields; thinned point process.	

\end{abstract}

\section{Introduction}

\begin{figure}
	\centering
	\includegraphics[width=\textwidth]{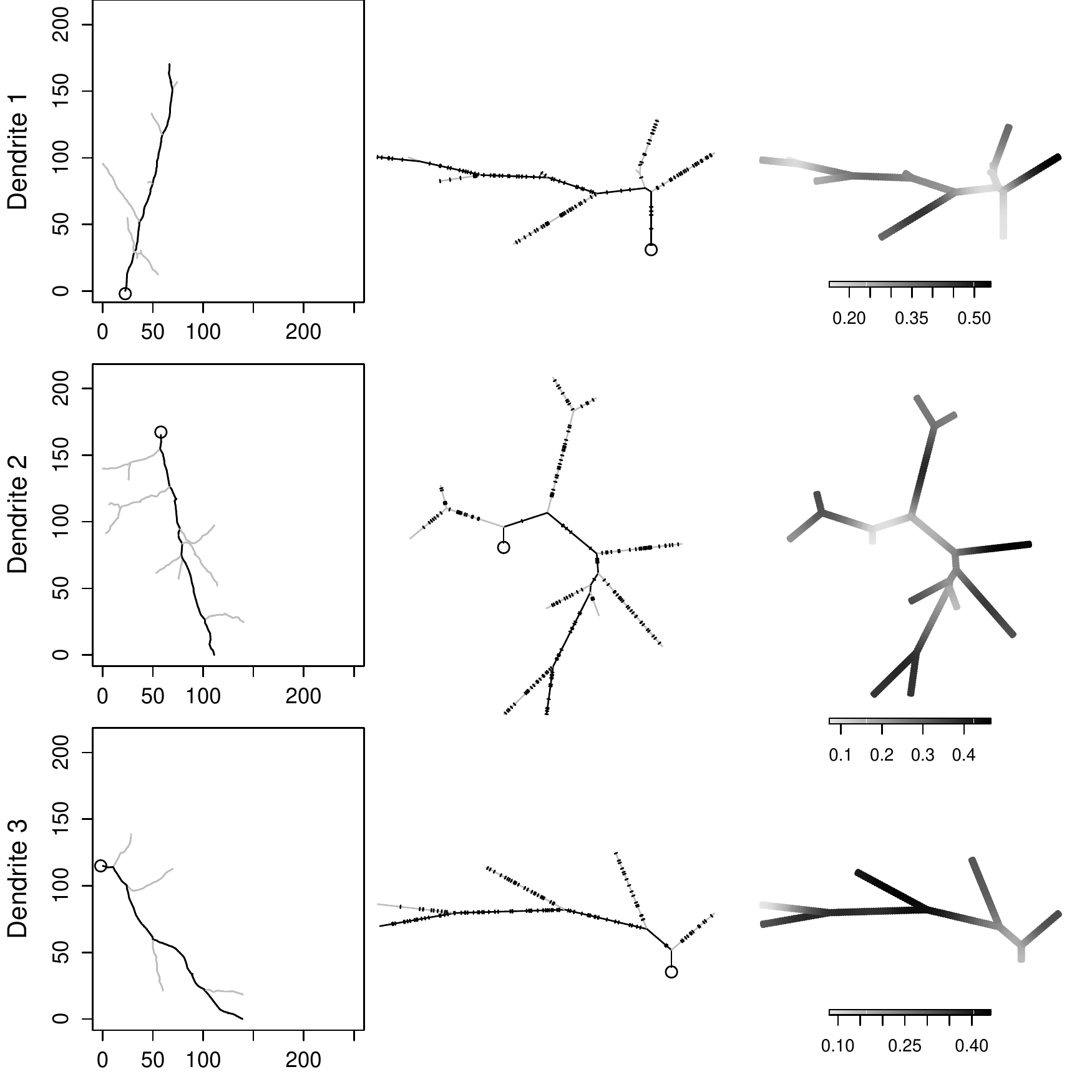}
	\caption{Spine data sets for dendrites 1 to  3 (from top to bottom),
		where each main branch is coloured black and the side branches
		grey and the $\circ$ marks the vertex closest to the dendrite's
		attachment to soma.  Left: projection of the original
		three-dimensional network onto a plane. Middle: spine locations on
		the simplified networks embedded in $\mathbb{R}^2$ so that
		distances are preserved; for details, see Section~\protect{\ref{sec:data_description}}. Right: non-parametric kernel intensity
		estimates; for details, see Section~\protect{\ref{sec:poisson_analysis}}.}
	\label{fig:data_1to3}
\end{figure}

\begin{figure}
	\centering
	\includegraphics[width=\textwidth]{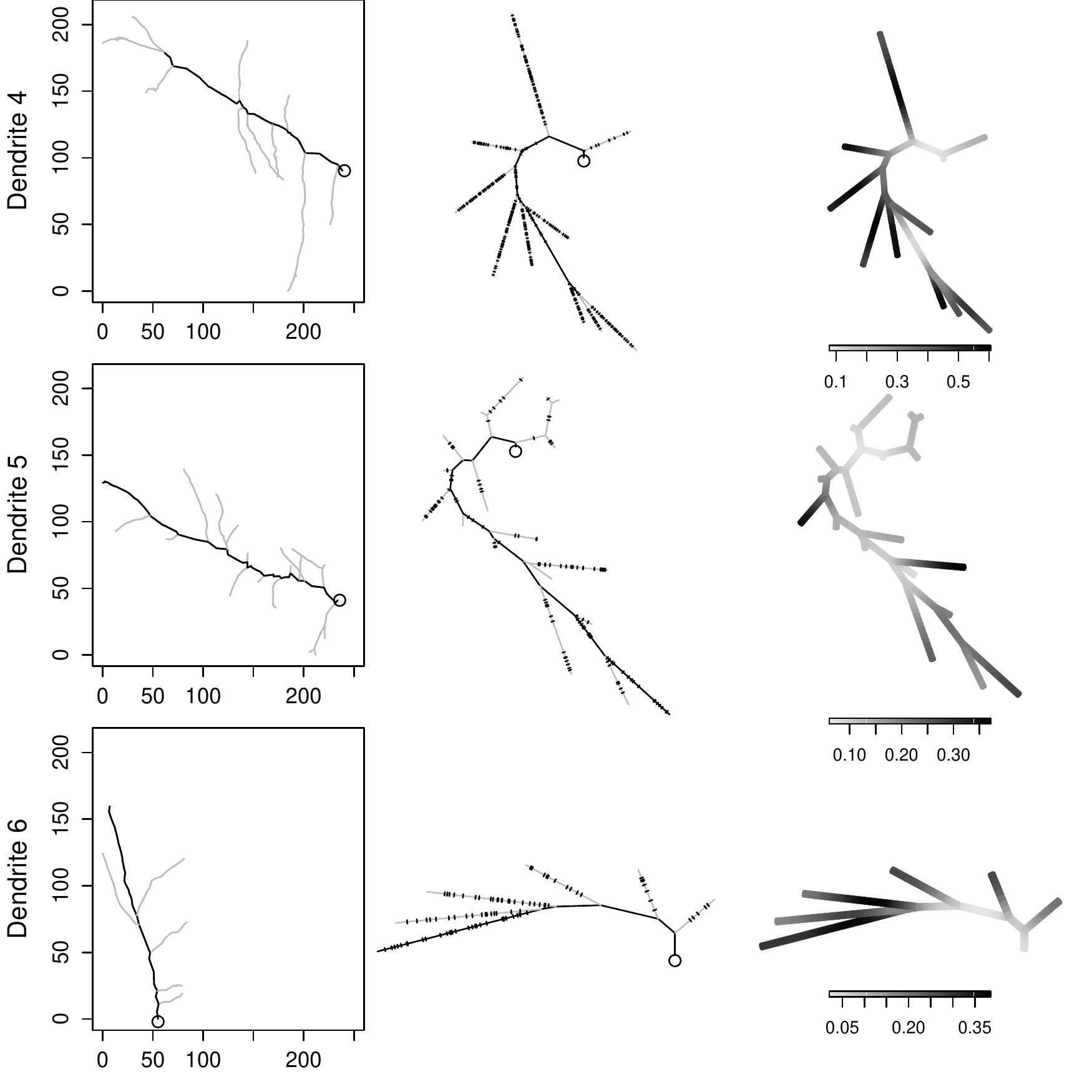}
	\caption{Spine data sets for dendrites  { 4 to 6} (from top to bottom),
		where each main branch is coloured black and the side branches
		grey and the $\circ$ marks the vertex closest to the dendrite's
		attachment to soma.  Left: projection of the original
		three-dimensional network onto a plane. Middle: spine locations on
		the simplified networks embedded in $\mathbb{R}^2$ so that
		distances are preserved; for details, see Section~\protect{\ref{sec:data_description}}. Right: non-parametric kernel intensity
		estimates; for details, see Section~\protect{\ref{sec:poisson_analysis}}.}
	\label{fig:data_4to6}
\end{figure}

Point patterns on linear networks arise in a broad range of fields, where the network for example represents roads, a river network, or a dendrite tree. This paper focuses on the latter type of data: the left panel in 
Figures~\ref{fig:data_1to3}--\ref{fig:data_4to6} shows six linear networks each representing a dendrite tree from a mouse neuron grown in vivo. On the dendrites small protrusions called spines are found that among other things help transmitting electrical signals to the soma. In neuroscience, the behaviour of spines is of interest as changes can be linked to changes in cognitive processes. The spine locations can be viewed as a point pattern on the dendrite tree and thus analysed using point process theory for linear networks. These six spine point pattern data sets will hereafter be referred to as the `spine data'.

Over the last two decades, methods for analysing point patterns on linear networks have been developed. 
Particularly, a network analogue of Ripley's $K$-function was first presented in \cite{okabe-yamada-01} and later { modified} and extended to the inhomogeneous case in \cite{ang-etal-12}. When defining the $K$-function, \cite{ang-etal-12} required that the underlying point process model fulfils an invariance property called second-order pseudo-stationarity \citep[an analogue to second-order intensity-reweighted stationarity as introduced in][]{baddeley-etal-00}. This property is fulfilled whenever the pair correlation function is isotropic, i.e.\ when it only depends on the shortest path distance. \cite{baddeley-etal-17} showed that certain constructions, e.g.\ special types of Cox point processes that lead to point processes in the Euclidean space with an isotropic pair correlation function rarely result in second-order pseudo-stationary point processes when adapted to linear networks. 
Even without the requirement of pseudo-stationarity, there are { currently} only a limited number of point process models available for linear networks. For point processes on directed acyclic linear networks, \cite{rasmussen-christensen-18} presented both regular and clustered models defined by a generalisation of the conditional intensity function for temporal point processes.
\cite{anderes-etal-17} supplied a list of valid isotropic covariance functions for connected linear networks 
that can be used to construct Cox point processes, particularly log Gaussian Cox processes \citep[LGCPs; see also][]{moller-98}. 

Only few studies use point process theory to analyse the behaviour of spines: 
treating the dendrite tree as a directed tree, \cite{rasmussen-christensen-18} analysed {the spine data for dendrite 4 (visualised in the first row of Figure~\ref{fig:data_4to6})}.
The distribution of spines (and their shape) have further been investigated using point process theory in \cite{jammalamadaka-etal-13} (testing a homogeneous Poisson process model) and \cite{baddeley-etal-14} (using multitype Poisson process models to account for the shape classification) for in vitro grown neurons. Based on the network $K$-function, \cite{jammalamadaka-etal-13} concluded that a homogeneous Poisson process model seems adequate to describe the spine locations. However, \cite{jammalamadaka-etal-13} also stated that their results for the in vitro setting are unlikely to hold in an in vivo setting. 

Instead of Poisson process models, this paper suggests a new class of Cox process models on a linear network. Such a model applies for an undirected graph and is not a LGCP, but its construction still exploits a Gaussian random field so that the covariance functions from \cite{anderes-etal-17} become useful. Moreover, seemingly for the first time in connection to point process model fitting on linear networks, we demonstrate the use of minimum contrast and composite likelihood estimation procedures. Finally, we introduce new empirical summary functions and demonstrate their usefulness for model checking.

{ Our new empirical summary functions are analogous to the empirical $F$-, $G$-, and $J$-functions in \cite{lieshout-11} for inhomogeneous point processes on a Euclidean space \citep[][introduced a version of $F$- and $G$-functions but for homogeneous point processes on a linear network]{okabe:sugihara:2012}. We also consider empirical intensity functions, and after completing the first version of our paper we became aware of later appearing papers on intensity estimation \citep{rakshit-etal-19} and summary statistics  \citep{cronie-etal-19} for inhomogeneous point processes on a linear network. In particular, \cite{cronie-etal-19} introduced other kinds of empirical $F$-, $G$-, and $J$-functions which may be preferred from ours since their functions account for the geometry of the linear network.}    

The paper is organised as follows. The spine data is described in more detail in Section~\ref{sec:data_description} along with the general notion of a linear network. In Section~\ref{sec:pp_on_linear_network} we discuss existing as well as our new summary functions for point processes on linear networks; these are used for analysing the spine locations in Section~\ref{sec:modellingspines}. We initially suggest to model the spine locations by an inhomogeneous Poisson process model in Section~\ref{sec:poisson_analysis}, but due to clustering between spines we propose in Section~\ref{sec:coxmodel} an inhomogeneous Cox process model. Lastly, we discuss in Section~\ref{sec:discussion} possible extensions and future research directions.


\section{Dendritic spine data}\label{sec:data_description}
The spine data origin from six apical dendrite trees corresponding to two neurons from each of three different mice,
with each mouse neuron grown in vivo. 
The numbering of the dendrites is as follows: dendrite 1 and 2 come from mouse no.\ 1; dendrite 4 and 5 from mouse no.\ 2; and dendrite 3 and 6 from mouse no.\ 3. The first and middle columns of Figures~\ref{fig:data_1to3}--\ref{fig:data_4to6} show different ways of viewing each dendrite as a linear network (more details are given below). 
Specifically, a linear network is a union $L = \bigcup_{i = 1}^N L_i$ of a finite number $N$ of line segments $L_i\subset \mathbb{R}^d$, $d \geq 2$, with finite length and intersecting only at the end points. A linear network may also be viewed as a graph consisting of a set of vertices and a set of weighted edges, where the edges coincide with the line segments $L_1, \dots, L_N$, the vertices correspond to the end points of these line segments, and the weight of an edge is the length of the corresponding line segment. 

Throughout this paper we assume that $L$ is a connected set and the distance between two points $u, v \in L$ is measured by the shortest path distance and  is denoted by $d_L(u, v)$. 
For the spine data, the linear network $L$ of a dendrite is a tree, meaning that there is only one path between any pair of points in $L$. 
Naturally, in other applications more complicated networks than a tree occur in which case we may need to take more care when choosing the distance metric $d_L$ (see Section~\ref{sec:discussion} for details).

The linear networks visualised in the left column of Figures~\ref{fig:data_1to3}--\ref{fig:data_4to6} are approximations of the underlying apical dendrite trees. The vertices of each of the linear networks are described by three-dimensional coordinates which represent a spine location or another point chosen to obtain the approximation. For each dendrite tree, we talk about two subsets: the main branch and the side branches. 
\textit{Main branch} refers to the tree's stem, while \textit{side branches} constitute the rest of the tree.
The left column of Figures~\ref{fig:data_1to3}--\ref{fig:data_4to6} shows which parts of the
approximations of the underlying apical dendrite trees
belong to the main branch and which to the side branches.

{ To utilise the functionalities of the \verb|R|-package \verb|spatstat| \citep{baddeley-etal-15}, each of the linear networks has been transformed into a simpler network embedded in $\mathbb{R}^2$ in the following way. First, we reposition any sequence of edges which are consecutively connected by vertices of degree two (such edges will either be contained in the main branch or in the same side branch) with a single edge/line segment whose weight is the sum of the old edge weights. This procedure straightens out `kinks' in any connected subset of the network containing only such line segments, as the angle between a pair of neighbouring edges becomes $180^\circ$.  Next, any line segments meeting at a vertex of degree higher than two can be repositioned such that the entire network is contained in a plane with no overlap between line segments (except at the end points). Thereby the network has naturally been embedded in $\mathbb{R}^2$. Note that this transformation relies on the fact that the linear network forms a tree, and that it preserves the distance between any pair of points in the network. Most importantly, the transformation allows us without any loss of information to consider 
the spine locations as a point pattern on the simplified and embedded linear network when in Section~\ref{sec:modellingspines} we are dealing with 
\begin{itemize}
\item
non-parametric kernel estimates of the intensity, since they will be unchanged under the performed transformation,
\item 
specific Poisson and Cox process models and statistical tools, since as we shall see
they will 
 not directly depend on the three-dimensional coordinates but only on 
 \begin{itemize}
 \item how many points in a spine point pattern data set will be on the main branch and on the side branches, respectively,
 \item the
distances between pairs of points.
\end{itemize}   
\end{itemize}
}

The simplified and embedded versions of the networks are shown in the middle column of Figures~\ref{fig:data_1to3}--\ref{fig:data_4to6} along with the spine locations. It is also shown which parts of the simplified embedded trees belong to the main branch and which to the side branches. 
In the following, $L = L_m \cup L_s$ refers to one of the six simplified and embedded linear networks in Figures~\ref{fig:data_1to3}--\ref{fig:data_4to6}, where $L_m$ is the main branch, and $L_s$ is the union of the side branches.  
Further, $n_{m}$ and $n_{s}$ denote the number of spines on $L_m$ and $L_s$, respectively. 
Lastly, we let $|B|$ denote the size of $B \subseteq L$ or more precisely the total length of the (partial) line segments constituting $B \subseteq L$; note that $|L| = |L_m| +|L_s|$. Table~\ref{tab:dendrite_summaries} summarises the number of spines and sizes for each dendrite tree.

\begin{table}
	\centering
	\caption{Number of spines, length, and intensity estimates for the main and side branches separately, where the intensity estimates are for the parametric model \eqref{eq:intensity} given in Section~\protect{\ref{sec:poisson_analysis}}.}
	\label{tab:dendrite_summaries}
	\begin{tabular}{c|cc|cc|cc}
		\toprule
		Dendrite & $n_m$ & {$n_s$} & $|L_{m}| $             & $|L_{s}|$                 & $\hat{\rho}_{m}$ & $\hat{\rho}_{s}$ \\ 
		\midrule
		1        & 51    & 72    & \SI{212}{\micro\meter} &  \SI{202}{\micro\meter} & 0.240            & 0.356            \\ 
		2        & 36    & 145   & \SI{204}{\micro\meter} &  \SI{430}{\micro\meter} & 0.176            & 0.337            \\ 
		3        & 69    & 63    & \SI{211}{\micro\meter} &  \SI{202}{\micro\meter} & 0.328            & 0.312            \\ 
		4        & 33    & 308   & \SI{225}{\micro\meter} &  \SI{652}{\micro\meter} & 0.134            & 0.477            \\  
		5        & 34    & 83    & \SI{286}{\micro\meter} &  \SI{450}{\micro\meter} & 0.119            & 0.184            \\  
		6        & 30    & 62    & \SI{178}{\micro\meter} &  \SI{250}{\micro\meter} & 0.168            & 0.248            \\  
		\bottomrule
	\end{tabular}
\end{table}

\section{Point processes on linear networks}\label{sec:pp_on_linear_network}
The six spine point pattern data sets are modelled as realisations of six point processes defined on the six dendrite trees.  
In general, by a point process $X$ on a linear network $L$ we mean a random finite subset of $L$; we use this generic notation throughout this paper.
In this section we consider summary functions useful for analysing point processes on linear networks, including the introduction of new empirical summary functions.

\subsection{Summary functions for first and second-order moment properties}
We assume that $X$ has intensity $\rho$, that is, for $B \subseteq L$, 
\begin{equation}\label{eq:def_intensity}
\E n(X \cap B) =  \int_B \rho(u) \, \di_L u < \infty,   
\end{equation} 
where $n(X \cap B)$ is the number of points from $X$ falling in $B$ and $\di_L$ denotes integration with respect to one-dimensional arc-length along $L$.  Intuitively, $\rho(u)\, \di_L u$ is the probability of $X$ having a point in an infinitesimal small subset of $L$ that contains $u$ and has size $\di_L u$. If the intensity  $\rho(\cdot) \equiv \rho$ is constant, we say that $X$ is homogeneous; otherwise $X$ is said to be inhomogeneous. In case of homogeneity, $\rho$ is the expected number of points per unit length. 

We also assume that $X$ has pair correlation function $g$, that is, for disjoint $A, B \subset L$,
\begin{equation*}
\E \{ n(X\cap A)n(X\cap B) \} = \int_A \int_B g(u, v) \rho(u) \rho(v) \, \di_L u \, \di_L v < \infty.
\end{equation*}
We can interpret $g(u, v) \rho(u) \rho(v) \, \di_L u \, \di_L v$ as the joint probability that two infinitesimal small regions around $u$ and $v$ of size $\di_L u$ and $\di_L u$, respectively, each contains a point from $X$.  

{ We say that the pair correlation function is isotropic if it only depends on the shortest path distance, that is, $g(u, v) =  g_0\{d_L(u, v)\}$ where $g_0$ is a non-negative function.}  
When $X$ has an isotropic pair correlation function, the (geometrically corrected network) $K$-function introduced by \cite{ang-etal-12} can be expressed as  
\begin{equation}\label{eq:def_Kfunction}
K(r) = \int_{0}^{r} g_0(s) \, \di s, \qquad r \geq 0.
\end{equation}
Alternatively, $K(r)$ may be written as an expectation with respect to a Palm distribution, see \citep[Theorem 3]{ang-etal-12}. If  the $K$-function or the pair correlation function is expressible on closed form, we can use a minimum contrast or composite likelihood procedure to estimate the model parameters; this is described further in Section~\ref{sec:estimation_procedure}. 

The simplest point process model is a Poisson process, which is characterised by that $n(X)$ follows a Poisson distribution with mean given by \eqref{eq:def_intensity} with $B = L$ and further that the points of $X$ conditioned on $n(X)$ are independent and identically distributed, with density proportional to $\rho$.  
For a Poisson process, $g \equiv 1$ and $K(r) =\nobreak r$.

\subsection{New empirical summary functions}\label{sec:modelcheck}

For estimating the pair correlation function and the $K$-function we follow \cite{ang-etal-12}. These empirical summary functions can be used in minimum contrast or composite likelihood estimation procedures as well as for model checking. Obviously, if the $K$-function or pair correlation function have been used to fit the model, neither should be used to check the adequacy of the model. 
Due to the shortage of summary functions for point processes on linear networks, we may let a simple visual comparison of the observed point pattern and simulations from the fitted model serve as a model check. It is needless to say that a more rigorous model checking would be preferred. 

Therefore, we now introduce three purely empirical summary functions. These are obtained by modifying the empirical $F$-, $G$-, and $J$-functions for inhomogeneous point patterns on a Euclidean space \citep[introduced by][]{lieshout-11} to linear networks. The modification simply consists of replacing the Euclidean space with the linear network, introducing the shortest path distance instead of the Euclidean distance, and adapting the notion of an eroded set to linear networks. The functions are then defined as follows.  
Assume that the intensity $\rho$ is known or has been estimated by $\hat{\rho}$ and that $\bar{\rho} = \inf_{u\in L} \tilde{\rho}(u) > 0$, where either $\tilde{\rho} = \rho$ or $\tilde{\rho} = \hat{\rho}$.
For $r \geq 0$, let $L_{\ominus r}$ consist of the points in $L$ with distance greater than $r$ to any vertex of $L$ with degree one. Furthermore, let $H\subset L$ be a finite `lattice'. { For a tree network the points of $H$ may be chosen equidistant, while on more complicated networks we can e.g.\ choose $H$ to be a collection of points that are equidistant within each line segment.}   For an observed point pattern $X = x$, the empirical summary functions $\hat{F}$, $\hat{G}$, and $\hat{J}$ are then defined for $r \geq 0$ by 
\begin{align}
\hat{F}(r) &= 1 - \frac{\sum_{v \in H \cap L_{\ominus r}}\prod_{u\in x :\, d_L(u, v) \leq r}\bigl\{1 - \frac{\bar{\rho}}{\tilde{\rho}(u)}\bigr\}}{  \# (H \cap L_{\ominus r})},  \label{eq:Fdef}\\
\hat{G}(r) &= 1 - \frac{\sum_{v \in x\cap L_{\ominus r}}\prod_{u\in x\backslash\{v\} :\, d_L(u, v) \leq r}\bigl\{1 - \frac{\bar{\rho}}{\tilde{\rho}(u)}\bigr\}}{  \# (x \cap L_{\ominus r})},\label{eq:Gdef}\\
\hat{J}(r) &= \frac{1 - \hat{G}(r)}{1- \hat{F}(r)},\label{eq:Jdef}
\end{align}
where we restrict attention to $r$-values small enough to ensure that $\# (H \cap L_{\ominus r}) > 0$ for $\hat F(r)$, $\#(x\cap L_{\ominus r})>0$ for $\hat G(r)$, and $\hat F(r)<1$ for $\hat J(r)$. { Further, $r$ should not be chosen larger than a reasonable amount of the network remains when considering $L_{\ominus r}$.}

In \cite{ang-etal-12}, the $K$-function and its empirical estimate include a factor that corrects for the network geometry, such that its shape can be compared for point patterns on different networks.  
As it was not obvious to us how to extend such a correction to $\hat{F}$, $\hat{G}$, and $\hat{J}$ in a meaningful way, our definitions in \eqref{eq:Fdef}--\eqref{eq:Jdef} do not correct for the network geometry. 
Further, we do not have any theoretical counterpart to $\hat{F}$, $\hat{G}$, and $\hat{J}$ and therefore their shapes alone can in general not be used to conclude anything about e.g.\ the presence of regularity or clustering. However, $\hat{F}$, $\hat{G}$, and $\hat{J}$ are still useful tools for providing a so-called global rank envelope; this is a confidence region for a given test function obtained from simulations under a fitted model \citep[for details, see][]{myllymaki-17}. In a global rank envelope procedure, the shape of the test function for the data is compared to that of the simulations and \cite{myllymaki-17} discussed how this provides a test and an interval with lower and upper bounds given by a liberal and a conservative $p$-value, respectively.  

\section{Modelling spine locations}\label{sec:modellingspines}
In this section each of the six data sets is analysed with the aim of finding a model that adequately describe the spine locations. 

\subsection{Fitting a Poisson process model}\label{sec:poisson_analysis}
The simplest model we can propose is a Poisson process. To investigate the behaviour of the spine intensity, we calculated the non-parametric intensity estimate suggested by \cite{mcswiggan-etal-16} using the \verb|density.lpp| function from the \verb|spatstat|-package {with Scott's rule of thumb modified to linear networks for choosing the bandwidth \citep{rakshit-etal-19}}; the resulting estimates are seen in the right panel of Figures~\ref{fig:data_1to3}--\ref{fig:data_4to6}, where the bandwidth is 18, 20, 18, 22, 23, 14 for dendrite $1, \dots, 6$, respectively. 
Note that the estimated spine intensity tends to be higher on the side branches than on the main branch (except perhaps for dendrite 3) and it seems plausible to assume a constant intensity value $\rho_m$ and $\rho_s$ on the main and side branches, respectively (with different $(\rho_m,\rho_s)$-values for the six dendrites). Therefore, 
recalling the notation in Section~\ref{sec:data_description}, we assume that
\begin{equation}\label{eq:intensity}
\rho(u) = \rho_m^{\mathbb{I}(u \in L_m)}\rho_s^{\mathbb{I}(u \in L_s)}, \qquad u\in L,
\end{equation}
where 
$\mathbb{I}(\cdot)$ denotes the indicator function. Naturally, we could also propose inhomogeneous Poisson process models involving other covariates (than $\mathbb{I}(u \in L_m)$ and $\mathbb{I}(u \in L_s)$) on the network. As no quantities beside the location of the spines have been recorded, the only covariates available are those that we can create within the network, such as distance to a vertex of degree higher than two. However, as we see no clear relation between such covariates and the intensity function we restrict ourselves to consider the Poisson process model specified by \eqref{eq:intensity}.  

The maximum likelihood estimates of the intensity parameters in \eqref{eq:intensity} are easily found and given by
\begin{equation}\label{eq:rhoest_poisson}
\hat{\rho}_m = \frac{n_m}{|L_m|}, \qquad \hat{\rho}_s =
\frac{n_s}{|L_s|},
\end{equation}
cf.\ the notation in Section~\ref{sec:data_description}.
These estimates are shown in Table~\ref{tab:dendrite_summaries}. 

\begin{figure}[htbp]
	\centering
	\includegraphics[width=0.8\textwidth]{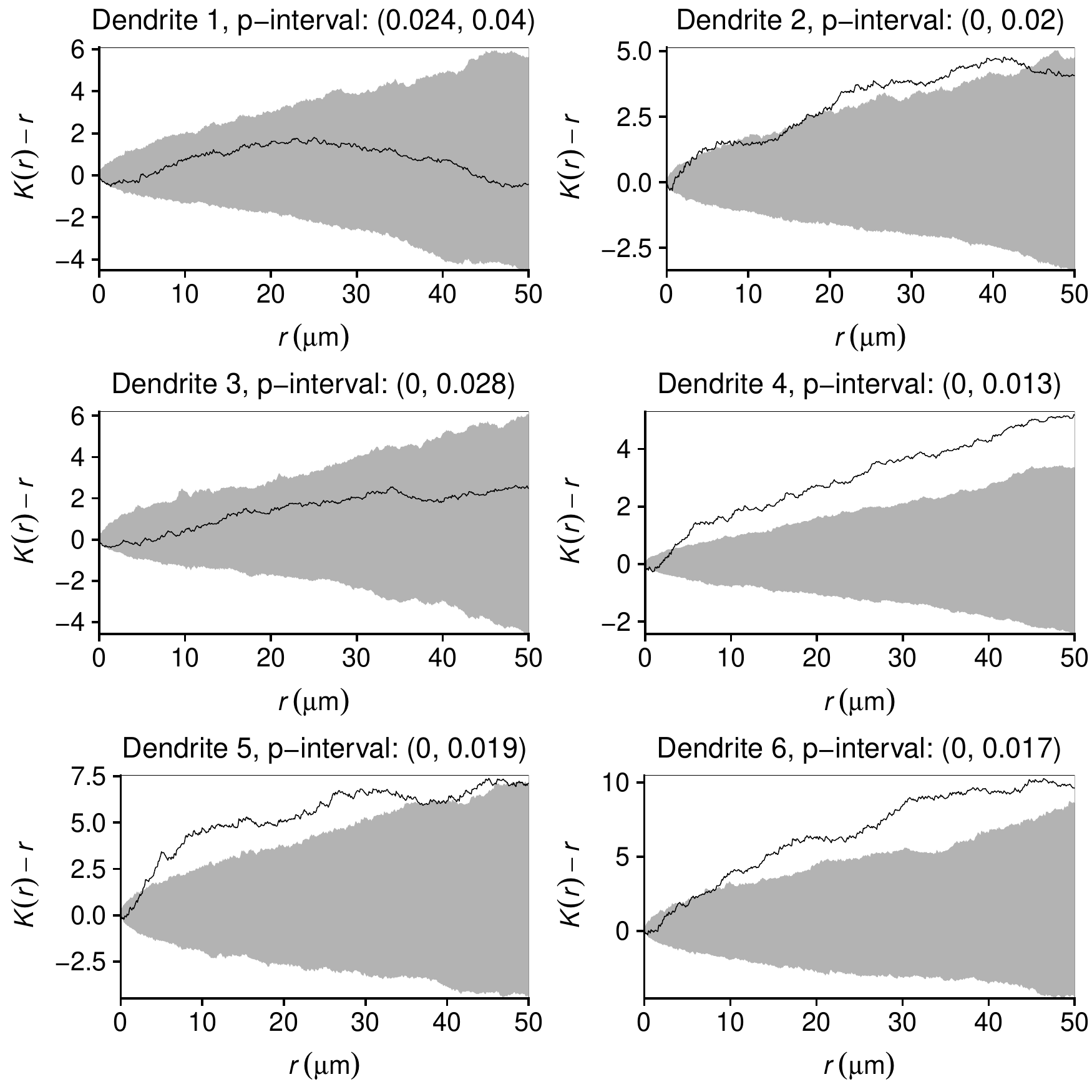}
	\caption{For each spine data set: the empirical $K$-function for the
		data minus the $K$-function for a Poisson process (solid line)
		along with $95 \%$ global rank envelopes (grey region) based on
		2499 simulations from the fitted inhomogeneous Poisson process
		model; $p$-intervals for each of the associated global rank
		envelope tests are also displayed.}
	\label{fig:Poisson_gre_K}
\end{figure}

To test whether the proposed inhomogeneous Poisson process model
adequately describes the spine locations, we performed global rank
envelope tests using $K$ as test function, cf.\
Section~\ref{sec:modelcheck}. Results from these tests are shown in
Figure~\ref{fig:Poisson_gre_K}. For all dendrites the conservative
$p$-value is below $3\%$, suggesting that the fitted Poisson process
models do not describe the spine locations adequately.  Specifically,
the empirical $K$-functions for dendrite 2, 4, 5, and 6 fall above the
global rank envelopes for certain $r$-values, indicating that the
spines tend to cluster at these distances. Further, for all six spine
data sets the empirical $K$-function falls below the global rank
envelope for small $r$-values, indicating a small-scale repulsion
between spines. For dendrite 1 and 3, this small-scale repulsion is
the only deviation from the Poisson process model revealed by the
global rank envelope test.  Disregarding the small distances
($r < \SI{1}{\micro\meter}$) for the global rank envelope test with
$K$ as test function, does not change the $p$-intervals significantly
for dendrite 2, 4, 5, and 6. For dendrite 1 and 3 on the other hand,
the $p$-intervals change from $(0.024, 0.040)$ to $(0.048, 0.060)$ and
from $(0, 0.019)$ to $(0.160, 0.168)$, respectively, giving (most
clearly for dendrite 3) no evidence against the proposed Poisson
process model. Global rank envelopes with a concatenation of
$\hat{F}$, $\hat{G}$, and $\hat{J}$ as test function, and where
distances less than $\SI{1}{\micro \meter}$ were disregarded, are
shown in Figure~\ref{fig:JFGPoisson} in Appendix~\ref{app:C}; these do
not provide any evidence against the Poisson process model for
dendrite 1 and 3 either.

As the physical scale of the spine data is quite small (the dendrites
range in size from \SI{412}{\micro \meter} to \SI{876}{\micro \meter},
cf.\ Table~\ref{tab:dendrite_summaries}) and as there is uncertainty
in the exact choice of the point representing a spine's location, we
must expect some degree of imprecision and therefore we may not want
to put too much value into the observed small-scale repulsion. In the
following we will not take the small-scale repulsion into account but
rather focus on modelling the large scale clustering.

\subsection{Fitting a Cox process model}\label{sec:coxmodel}
In addition to inhomogeneity in the location of spines, Figures~\ref{fig:data_1to3}--\ref{fig:data_4to6} also shows a tendency to large bare areas where no spines occur. To model this behaviour we considered a point process model introduced by \cite{lavancier-moeller-16} for the Euclidean space, which is easily adapted to a linear network $L$. The point process is constructed as a thinning of a Poisson process $Y$ on $L$ with intensity function $\rho_Y$, where the retention probabilities are determined by a random field $\Pi = \{\Pi(u) : u \in L\}$ that may be spatially correlated. That is, the point process is given by $X = \{u \in Y: \Pi(u) \geq R(u)\}$, where $R = \{R(u) : u \in L\}$ consist of independent uniform random variables on $[0, 1]$, and where $Y$, $\Pi$, and $R$ are independent. Thus $X$ is a Cox point process driven by the random field $\Lambda = \{\rho_Y(u)\Pi(u) : u \in L\}$. We let
\begin{equation}\label{eq:Pi}
\Pi(u) = \exp\Bigl\{- \frac{\sigma^2}{2}\sum_{j =
	1}^{k}Z_j^2(u)\Big\}, \qquad u\in L,
\end{equation}
where $k \in \{1, 2, \dots \}$ and $\sigma^2 > 0$ are parameters, and $Z_1, \dots, Z_k$ are IID zero-mean unit-variance Gaussian random fields with correlation function $c$. If $c(u, v) = c_0\{d_L(u, v)\}$ depends only on the shortest path distance, we say that $c$ is isotropic; see \cite{anderes-etal-17} for a list of isotropic correlation functions for linear networks. For the spine data, we considered the exponential correlation function, that is,
\begin{equation}\label{eq:expcov}
c(u, v) = \exp\{-\beta d_L(u, v)\}, \qquad u, v \in L,
\end{equation}
which is a valid correlation function for any parameter value $\beta > 0$ and any tree network but not necessarily for other kinds of linear networks (see Section~\ref{sec:discussion} for a comment on this). 

We have that $\E \Pi(u) = (1 + \sigma^2)^{-k/2}$ and 
\begin{equation*}
\E \{ \Pi(u)\Pi(v)\} = \{(1 + \sigma^2)^2 - (\sigma^2)^2 c(u, v)^2\}^{-k/2},
\end{equation*}
implying that 
$X$ has intensity  
\begin{equation}\label{eq:rhoX}
\rho(u) 
= (1 + \sigma^2)^{-k/2} \rho_Y(u), \qquad u\in L, 
\end{equation}
and pair correlation function
\begin{equation}\label{eq:gX}
g(u, v)
=\left\{\frac{(1 + \sigma^2)^2}{(1 + \sigma^2)^2 - (\sigma^2)^2 c(u, v)^2}\right\}^{\mathrlap{k/2}}, \qquad u, v\in L, \, u\neq v.
\end{equation}
If $c$ is isotropic, then $g$ is isotropic and the $K$-function can be expressed by \eqref{eq:def_Kfunction}. Closed form expressions of the $K$-function are given in Appendix~\ref{app:A} for $c$ equal to the exponential correlation function and $k = 1, \dots, 5$. 

Note that for each $u \in L$, $\sigma^2\sum_{j = 1}^{k}Z_j^2(u)$ in \eqref{eq:Pi} follows a $\sigma^2 \chi^2$-distribution with $k$ degrees of freedom.  
That is, the skewness decreases as $k$ increases, while the range is stretched/compressed depending on the value of $\sigma^2$. The larger $\sigma^2$ is, the more $Y$ is thinned to obtain $X$ and also the more variation in the thinning probabilities. The pair correlation function in \eqref{eq:gX} is an increasing function of both $\sigma^2$ and $k$. 
When $c$ is given by \eqref{eq:expcov}, $\beta$ controls the
correlation of the retention probabilities: the smaller~$\beta$, the
longer range of correlation in $\Pi$ and thus larger coherent
bare/populated areas in $X$. Finally, the pair correlation function
decreases towards 1, when $\beta$ increases.

For the spine data, we still want to model the intensity function by \eqref{eq:intensity}, which by \eqref{eq:rhoX} requires $Y$ to have a similar intensity structure, that is, 
\begin{equation}\label{eq:rhoY}
\rho_Y(u) = \rho_{Y, m}^{\mathbb{I}(u \in L_m)}\rho_{Y, s}^{\mathbb{I}(u \in L_s)}, \qquad u \in L,
\end{equation}
for non-negative parameters $\rho_{Y, m}$ and $\rho_{Y, s}$. 

\subsubsection{Simulation}
To perform model checking with global rank envelopes or to carry out simulation studies, we need to be able to simulate point patterns from the model of interest. Fortunately, it is straightforward to simulate a point pattern on $L$ from the proposed Cox process model by the following three steps:
\begin{enumerate}
	\item[(a)] Simulate a discretised version of the random field $\Pi$ by first simulating the independent Gaussian random fields $Z_j$, $j = 1, \dots, k$, at chosen grid locations along the network and then transforming the random fields according to \eqref{eq:Pi} to obtain the retention probabilities.
	\item[(b)] Simulate a point pattern $y$ from a Poisson process on $L$ with intensity $\rho_Y$.
	\item[(c)] Thin $y$ using the retention probabilities simulated in (a).  
\end{enumerate}

\subsubsection{Estimation procedure} \label{sec:estimation_procedure}
In the following we describe a procedure for estimating the model parameters of the proposed Cox process model where. For specificity we consider the case where $c$ is given by \eqref{eq:expcov} and where $\rho_Y$ is given by \eqref{eq:rhoY}. 

To begin we assume that $k$ is known, whereas the remaining parameters are estimated through a two-step procedure \citep{waagepetersen-07, waagepetersen-guan-09}. In short, we first estimate $(\rho_m, \rho_s)$ and then plug in these estimates in a second-order procedure where $(\sigma^2, \beta)$ is estimated. Lastly, an estimate of $(\rho_{Y, m}, \rho_{Y, s})$ can be found by using \eqref{eq:rhoX}.

First, to estimate $(\rho_{m}, \rho_{s})$ we use the first order composite likelihood  \citep{waagepetersen-07} which simply corresponds to a Poisson likelihood yielding the estimates in \eqref{eq:rhoest_poisson}.

Second, as we know explicit formulas for the pair correlation and $K$-function, we can estimate $(\sigma^2, \beta)$ using a minimum contrast procedure \citep{guan-09, diggle-14} or a second-order composite likelihood approach \citep{waagepetersen-07, lavancier-etal-18}. The latter is not considered here but described in Appendix~\ref{app:B}, where results from a simulation study comparing the two approaches also can be found. The simulation study suggests that the minimum contrast procedure provides {more meaningful estimates with less bias and variance} than the second-order composite likelihood. 

For a chosen summary function   $T_{(\sigma^2, \beta)}$ which depends on $(\sigma^2, \beta)$, the minimum contrast estimate of $(\sigma^2, \beta)$ is given by
\begin{equation}\label{eq:mincon_est}
(\hat{\sigma}^2, \hat{\beta}) = \underset{(\sigma^2, \beta)}{\arg\min}  \int_{r_l}^{r_u} \bigl\{\hat{T}(r)^p - T_{(\sigma^2, \beta)}(r)^p\bigr\}^2 \, \di r,
\end{equation}
where $0 \leq r_l < r_u$ and $p > 0$ are user-specified tuning
parameters, and $\hat T$ is an empirical estimate of the summary
function. In our case, $\smash{T_{(\sigma^2, \beta)}}$ is given by $K$
or~$g_0$, and $\hat T$ is given as in \cite{ang-etal-12}. A frequently
seen choice is $r_l = 0$, while general recommendations of $r_u$ and
$p$ can be found in \cite{guan-09} and \cite{diggle-14} for point
patterns on the Euclidean space.

The minimum contrast procedure can easily be extended to include estimation of $k$ too, but a simulation study indicated that it may be difficult to estimate $\sigma^2$ and $k$ simultaneously as an increase in $k$ seemingly can be balanced out by an increase in $\sigma^2$. 
In practice we may therefore simply make a choice of $k$; for simplicity we chose $k = 1$ in the following. 
Note that  for both $T = K$ and $T = g_0$ the estimate $\hat{T}$ in \eqref{eq:mincon_est} depends on the intensity \citep{ang-etal-12}; here we simply plug-in the estimated intensity obtained in the first step of the estimation procedure. 

A drawback of using $T = g_{0}$ is the need of choosing a bandwidth for the non-parametric kernel estimate $\hat{g}_{0}$ presented in \cite{ang-etal-12}. However, the simulation study in  Appendix~\ref{app:B} suggests that using $T = g_0$ with the default bandwidth and kernel from the \verb|spatstat|-package {generally yield estimates with lower bias and variance} than $T = K$ when fitting the proposed Cox process model. This is consistent with results from a simulation study in \cite{lavancier-moeller-16} for point processes on a Euclidean space. 

In the simulation study found in Appendix~\ref{app:B}, we also investigated how different choices of $r_l$, $r_u$, and $p$ affect the estimates of $\sigma^2$ and $\beta$ given by \eqref{eq:mincon_est}.
We observed that the choice of $r_u$ often is a matter of trade-off between bias and variance: a large value of $r_u$ may entail a large bias, while a small $r_u$ often leads to a greater variance of the estimates. { The best choice of $r_u$ also} seems to be quite depending on what the true underlying model parameters are. For example, a larger range of correlation in the retention probabilities, that is, a smaller value of $\beta$, requires a larger $r_u$. Naturally we should also take the size of the network into consideration when choosing $r_u$. 
In the simulation study we found that $r_l = 0$ gives the best estimates in terms of {bias and variance}, and that $p = 1$ or $p = 1/2$ behave equally well for $T = g_0$, while $p = 1/4$ is preferred over $p = 1/2$ for $T = K$. 
For parameter values yielding models close to the Poisson process model, that is, when $\sigma^2$ is close to zero or $\beta$ is large, the estimation procedures were not very successful regardless of the tuning parameters. This does not come as a surprise as many combinations of  $\sigma^2$- and $\beta$-values yield similar Poisson processes. 
Lastly, the estimation procedure seems quite stable with respect to the choice of start parameter values for the optimisation algorithm (\verb|optim| in \verb|R|) used to minimize \eqref{eq:mincon_est}.

\subsubsection{Model fit and model check}\label{sec:coxfitandcheck}
The Cox process model was fitted to each of the spine data sets using
the two-step procedure with $k = 1$ fixed, cf.\
Section~\ref{sec:estimation_procedure}. For the minimum contrast
procedure we let $T = g_0$, $p = 1$, and $r_l = 0$ in accordance with
the simulation results discussed in
Section~\ref{sec:estimation_procedure}. Further, we initially let
$r_u = 15$ and obtained a set of initial parameter estimates for each
data set. Then we performed a small simulation study based on 500
simulations from the initially fitted models to investigate which of
$r_u = 15$, $30$, $50$ results in the best estimates (in terms of bias
and variance) for these specific models. For dendrite 1, 3, and 6, the
initially fitted models are close to the Poisson process case, and as
a consequence the model parameters are hard to estimate regardless of
the choice of $r_u$. However, for dendrite 2, 4, and 5, it seems that
$r_u = 50$ is the best choice. Using $r_u = 50$ for all six data sets,
we obtained the parameter estimates in Table~\ref{tab:cox_par_est}.
The fitted model for dendrite 3 is practically a Poisson process model
(in consistency with the conclusions made in
Section~\ref{sec:poisson_analysis}) while the remaining fitted models
are not.  In Figure~\ref{fig:RF_simulation} in Appendix~\ref{app:C},
one simulation from each of the fitted random fields $\Pi$ is shown to
illustrate the behaviour of the retention probabilities. For example,
$\hat{\sigma}^2$ is considerably larger for dendrite~5 than
dendrite~1, resulting in more fluctuating retention probabilities.

\begin{table}[htbp]
	\centering
	\caption{Estimates of $\rho_{Y, m}$, $\rho_{Y, s}$, $\sigma^2$, and $\beta$ for each spine data set.}
	\label{tab:cox_par_est}
	\begin{tabular}{c !\qquad c !\quad c !\quad c !\quad c}
		\toprule
		Dendrite & $\hat{\rho}_{Y, m}$ & $\hat{\rho}_{Y, s}$  & $\hat{\sigma}^2$ & $\hat{\beta}$ \\ 
		\midrule
		1 & $0.312$ & $0.463$ & $0.686$  & $0.037$ \\ 
		2 & $0.275$ & $0.525$ & $1.427$ & $0.020$  \\ 
		3 & $0.328$ & $0.312$ & $5.170 \times 10^{-8}$  & $30.178$  \\ 
		4 & $0.197$ & $0.701$ & $1.159$ & $0.010$  \\ 
		5 & $0.266$ & $0.413$ & $4.023$ & $0.030$  \\ 
		6 & $0.322$ & $0.474$ & $2.662$ & $0.013$  \\
		\bottomrule
	\end{tabular}
\end{table} 

\begin{figure}
	\centering
	\includegraphics[width=\textwidth]{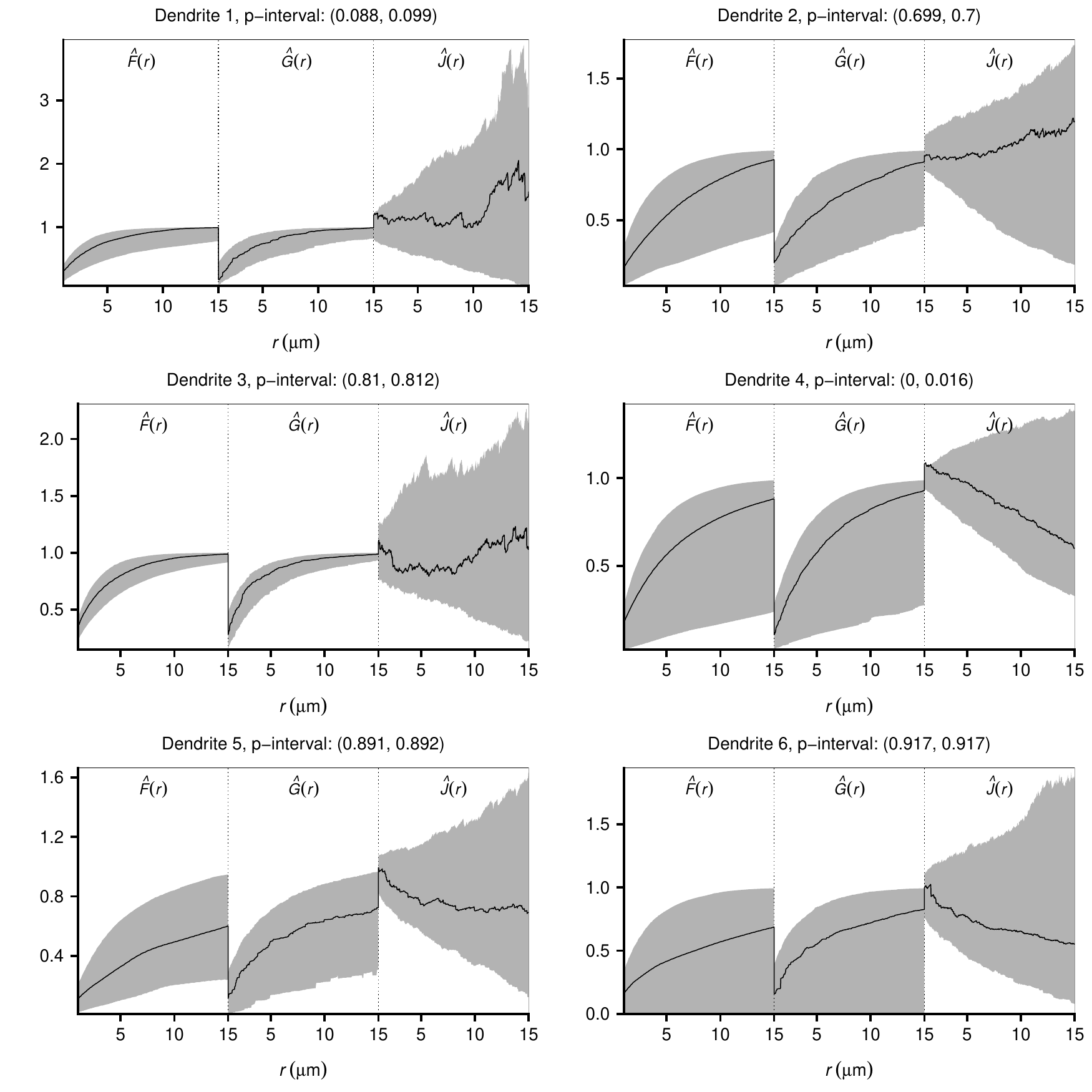}
	\caption{For each spine data set: the concatenation of $\hat{F}$,
		$\hat{G}$, and $\hat{J}$ for the spine locations (black solid
		line) along with $95\%$ global rank envelopes (grey region) based
		on 2499 simulations from the fitted Cox process model;
		$p$-intervals for each of the associated global rank envelope
		tests are also displayed. Here $r$-values less than $\SI{1}{\micro \meter}$ are disregarded.}
	\label{fig:JFGenv_Cox}
\end{figure}

As discussed in Section~\ref{sec:modelcheck}, for the statistical analyses of point patterns on linear networks there is only a limited number of options for model checking, especially when the $K$-function or the pair correlation function have already been used to estimate the model parameters. One simple option is to look at simulations from the fitted model as in Figures~\ref{fig:neuron1_simulations}--\ref{fig:neuron6_simulations} in Appendix~\ref{app:C}. Comparing these simulations visually to the observed point patterns, {it seems that the simulations mimic the behaviour of larger empty areas without spines seen in the data}. For a more rigorous model checking, we performed global rank envelope tests with a concatenation of $\hat{F}$, $\hat{G}$, and $\hat{J}$ as test function, where distances less than $ \SI{1}{\micro \meter}$ were disregarded as discussed in Section~\ref{sec:poisson_analysis}; results are shown in Figure~\ref{fig:JFGenv_Cox}. Except from dendrite 4, where the data curve for $\hat{J}$ falls below the global rank envelope for some of the smaller $r$-values, the tests do not reveal any evidence against the fitted models. However, for some of the dendrites (especially dendrite 6) the global rank envelopes for the part concerning $\hat{F}$ and $\hat{G}$ are very wide due to a large variance in the number of points. 

\section{Discussion}\label{sec:discussion}
{ Disregarding the possible small-scale repulsion discussed in Section~\ref{sec:poisson_analysis}, the fitted inhomogeneous Poisson models seemed adequate for dendrites 1 and 3, while the Cox model proposed in Section~\ref{sec:coxmodel} is preferred for the remaining dendrites.  
To incorporate the small-scale repulsion in the Cox model}, the inhomogeneous Poisson process $Y$ used to build the Cox process could be replaced by an inhomogeneous and repulsive point process. The simplest case may be to use a dependent thinning as in a Mat\'ern hard core process of type I \citep{matern:1960, matern:1986}: let $\tilde{Y}$ be an inhomogeneous Poisson process (with constant intensity on the main branch respectively the side branches), and let 
\[Y = \{u \in \tilde{Y} : d_L(u, v) > h \text{ for all } v\in \tilde{Y}\backslash\{u \}\},\] 
where $h > 0$ is a hard core parameter; that is, a point in $\tilde{Y}$ is included in $Y$ if and only if no other point in $\tilde{Y}$ is within distance $h$. 
However, it is doubtful whether an expression for the $K$-function or the pair correlation function can be found for such a Cox process model, posing new challenges with respect to parameter estimation. 

To avoid using the rather ad hoc created summary functions $\hat{F}$, $\hat{G}$, and $\hat{J}$, the summary functions proposed by \cite{cronie-etal-19} should be used instead. {These were first introduced after the preparation of this paper, but global rank envelope tests based on a concatenation hereof can be found in Figure~\ref{fig:JFGCoxCronie} in Appendix~\ref{app:C} for the fitted Cox model. 	
In our case, conclusions based on global rank envelopes using a concatenation of the $\hat{F}$, $\hat{G}$, and $\hat{J}$-functions presented in Section~\ref{sec:modelcheck} are consistent with those using the summary functions from \cite{cronie-etal-19}: except for dendrite 4, the observed summary functions stay strictly inside the 95\% global rank envelopes.}
	
\cite{rakshit-etal-17} discussed the importance of how distance is measured when analysing point patterns on a linear network and they generalised the $K$-function to allow the use of any distance metric. In fact, following \cite{anderes-etal-17} all methods as well as Poisson and Cox process models in this paper immediately apply for more general linear networks, called graphs with Euclidean edges, when the correlation function $c$ is isotropic with respect to the shortest path distance as well as another metric called the resistance metric. For the dendrite networks or any other tree network, the resistance metric is equivalent to the shortest path distance.
\cite{anderes-etal-17} showed that correlation functions that are isotropic with respect to the shortest path distance only are guaranteed to be valid for a small class of linear networks, whereas they are valid for any linear network when considering the resistance metric instead. Thus, depending on the network, it may be preferable to consider the resistance metric over the shortest path distance when specifying a correlation function. \cite{anderes-etal-17} provided a list of valid isotropic covariance functions for graphs with Euclidean edges. 

\appendix 	
\counterwithin{figure}{section}
\section{Expressions for $K$}\label{app:A}
For the Cox point process presented in Section~\ref{sec:coxmodel} with $c$ equal to the exponential correlation function in \eqref{eq:expcov}, the $K$-function is 
\begin{align}
K(r) = \int_{0}^{r} \left\{1 - \frac{\exp(-2\beta t)}{\left(1+1/\sigma^2\right)^2}\right\}^{-k/2} \, \di t.
\end{align}
Let $\alpha = \left(1 + 1/\sigma^2\right)^{-2}$, then $K(r)$ is given  by 
\begin{align*}
\frac{1}{\beta}\left\{\log\left(\sqrt{\mathrm{e}^{2\beta r} - \alpha} + \mathrm{e}^{\beta r}\right) - \log\left(\sqrt{1 - \alpha} + 1\right)\right\}
\end{align*}
if $k = 1$,
\begin{align*}
\frac{1}{2\beta} \log\left(\frac{\mathrm{e}^{2\beta r} - \alpha}{1 - \alpha}\right)
\end{align*} 
if $k = 2$,
\begin{align*}
\frac{1}{\beta}\biggl\{
&\frac{\sqrt{\mathrm{e}^{2\beta r} - \alpha} \log\bigl( \sqrt{\mathrm{e}^{2\beta r} - \alpha} + \mathrm{e}^{\beta r} \bigr) - \mathrm{e}^{\beta r} }{\mathrm{e}^{\beta r}\sqrt{1- \alpha \mathrm{e}^{-2\beta r}}} +\frac{1 - \sqrt{1 - \alpha}\log\bigl(\sqrt{1 - \alpha} + 1\bigr)}{\sqrt{1 - \alpha}}
\biggr\}
\end{align*}
if $k = 3$,
\begin{align*}
\frac{1}{2\beta}\Biggl\{ \frac{\alpha}{\alpha-\mathrm{e}^{2\beta r}} - \frac{\alpha}{\alpha - 1} + \log\Bigl( \frac{\mathrm{e}^{2\beta r} - \alpha}{1 - \alpha} \Bigr) \Biggr\}
\end{align*}
if $k = 4$,
and
\begin{align*}
\frac{1}{\beta} 
\biggl\{  
\frac{\alpha + \mathrm{e}^{-\beta r}(\mathrm{e}^{2\beta r} - \alpha)^{3/2}\log (\sqrt{\mathrm{e}^{2\beta r} - \alpha} + \mathrm{e}^{\beta r}) - \frac{4}{3}\mathrm{e}^{2\beta r}}{\sqrt{1 - \alpha \mathrm{e}^{-2\beta r}}(\mathrm{e}^{2\beta r} - \alpha)}-\log(\sqrt{1 - \alpha} + 1) + \frac{\frac{4}{3}-\alpha}{(1-\alpha)^{3/2}}
\biggr\}
\end{align*}
if $k = 5$.

\section{Simulation study concerning estimation procedure}\label{app:B}
\subsection{Second order composite likelihood}
In Section~\ref{sec:coxmodel} we fitted the parameters of the Cox process models using a two step procedure involving a minimum contrast procedure for estimating $\sigma^2$ and $\beta$. Another option is to consider a second order composite likelihood approach \citep[adapted from][to a point pattern on a linear network]{waagepetersen-07}. 
That is, for an observed point pattern $x \subset L$, the maximum composite likelihood estimate is obtained by maximizing the log composite likelihood
\begin{equation*}\label{eq:CL2_RW}
\begin{aligned}
CL(\sigma^2, \beta) = 
&\sum_{u, v \in x}^{\neq} w(u, v)\log\{\hat{\rho}(u)\hat{\rho}(v)g(u, v)\} - \int_L \int_L w(u, v)\hat{\rho}(u)\hat{\rho}(v)g(u, v) \di_L u \, \di_L v,
\end{aligned}
\end{equation*} 
where $w$ is a weight function and $\neq$ over the summation sign means that $u \neq v$. We can for example let 
\begin{align}\label{eq:indweight_nonadap}
w(u, v) = \mathbb{I}(d_L(u, v) \leq r_0)
\end{align} 
for some user specified value $r_0$.
To estimate $(\sigma^2, \beta)$, we either directly maximise \eqref{eq:CL2_RW} or alternatively solve the associated estimating equation obtained by setting the score equal to 0. The score function is in our set-up given by
\begin{equation}\label{eq:score_RW}
\begin{aligned}
\nabla CL(&\sigma^2, \beta) \\
&= \sum_{u, v \in x}^{\neq} w(u, v)\frac{\nabla g(u, v)}{g(u, v)} - \int_L \int_L w(u, v)\hat{\rho}(u)\hat{\rho}(v)\nabla g(u, v) \, \di_L u \, \di_L v.  
\end{aligned}
\end{equation}
To improve the composite likelihood estimation procedure, 
\cite{lavancier-etal-18} 
suggested an adaptive version of \eqref{eq:score_RW} where the weight function $w$ depends on the model parameters. We may for example let $w$ be the indicator function given by 
\begin{align}\label{eq:indweight_adap}
w(u, v) = \mathbb{I}\biggl(\frac{|g(u, v) - 1|}{M(u, v)} > \epsilon\biggr), 
\end{align}
where $M(u, v) = \max_{s \in \{u, v\}} \lvert g(s, s) - 1 \lvert$ and $\epsilon \in (0, 1)$ is a small user-specified number, e.g.\ $\epsilon = 0.01$ or $\epsilon = 0.05$. Note that for an isotropic correlation function $g(u, v) = g_0\{d_L(u, v)\}$, we have $M(u, v) = |g_0(0) - 1|$. Another weight function suggested in 
\cite{lavancier-etal-18} 
is 
\begin{align}\label{eq:expweight_adap}
w(u, v) 
= \begin{cases}
\exp\left[{1 / \{h(u, v)^2-1\}}\right] &\text{ for } -1 \leq h(u, v) \leq 1,\\
0 				  &\text{ else,}
\end{cases} 
\end{align}
where $h(u, v) = \epsilon M(u, v)/\{g(u, v) - 1\}$.

For approximating the double integral in \eqref{eq:score_RW} (or in the adapted version), note that this is of the form $\int_L\int_L f(u, v) \, \di_L u \, \di_L v$. We split up the integration area into the line segments constituting $L$, that is,
\begin{align*}
\int_L\int_L f(u, v) \, \di_L u \, \di_L v = \sum_{i, j}\int_{L_i}\int_{L_j} f(u, v) \, \di_L u \, \di_L v.
\end{align*}
Note that $\hat{\rho}(\cdot)$ for the spine data is constant on any line segment $L_i$ (as the line segment is either fully contained in $L_m$ or $L_s$). Further, if $L$ is a tree, $w$ is given by \eqref{eq:indweight_nonadap}, \eqref{eq:indweight_adap}, or \eqref{eq:expweight_adap}, and $g$ is isotropic, then $f(u, v) = f_0\{d_L(u, v)\}$ depends only on distance; this will ease the approximation of the integral:
\begin{align*}
\int_{L_i}\int_{L_j} f(u, v) \, \di_L u \, \di_L v
&=  
\begin{cases} 
\int_0^{|L_i|}\int_0^{|L_j|} f_0(d_{i, j} + x + y) \, \di x \, \di y \quad &\text{if } i \neq j,\\[0.5em]
\int_0^{|L_i|}\int_0^{|L_j|} f_0(|x - y|) \, \di x \, \di y \quad &\text{if } i = j,\\ 
\end{cases} 
\end{align*} 
where $d_{i, j} = \min_{u \in L_i, v \in L_j} d_L(u, v)$. Each of these integrals can then be approximated by Monte Carlo integration using uniform variables on $[0, |L_i|]$ and $[0, |L_j|]$.

\subsection{Simulation study}

In the following we describe and summarise results from a simulation study investigating how well $\sigma^2$ and $\beta$ are estimated using either the minimum contrast procedure with $T = K$ or $T = g_0$ (as described in Section \ref{sec:estimation_procedure}) or the adaptive composite likelihood procedure using \eqref{eq:indweight_adap} or \eqref{eq:expweight_adap}. We considered the network for dendrite 4 and simulated from the Cox process described in Section~\ref{sec:coxmodel} with different parameter values given in Table~\ref{tab:simstudypar} and when $k = 1$ was fixed.

For the minimum contrast procedure we investigated different values of the tuning parameters $r_u$, $r_l$, and $p$ as well as different start parameters for the optimisation algorithm, see Table~\ref{tab:simstudypar}. Here run no.\ 1 is the reference run from which one (or two) model or tuning parameters are changed at the time.   
For run no.\ 1, we chose $\sigma^2 = 5$ and $\beta = 0.1$ resulting in a model rather far away from the case of a Poisson process. Note that decreasing $\beta$ will increase the range of correlation in the thinning probability, whilst increasing $\sigma^2$ will increase the probability of thinning. Thus a small $\beta$ and a large $\sigma^2$ yield a model very different from the Poisson process. For each choice of model parameters we simulated 500 point patterns and estimated $(\sigma^2, \beta)$ using minimum contrast and for a few selected runs we also estimated $(\sigma^2, \beta)$ using the adaptive composite likelihood method.

For the adaptive composite likelihood method the integral in \eqref{eq:score_RW} was approximated using $10^6$ simulations. Estimates of $(\sigma^2, \beta)$ were found by minimising the length of the score over a $100 \times 100$ grid centred around the true values of $\sigma^2$ and $\beta$. The finer and broader grid, the better, but as a $100\times 100$ grid was already quite time consuming we settled with that. 
\begin{table}
	\centering
	\begin{tabular}{c|c|c|c|c|c|c|c|c}
		Run no. & $\sigma^2$ & $\beta$ & $\rho_{Y, m}$ & $\rho_{Y, s}$ & $p$ for MCE-$g$ (MCE-$K$) & $r_l$ & $r_u$  & $(\sigma^{2}_*, \beta_*)$ \\ \hline 
		1  & 5 & 0.1 & 0.8  & 1.2 & 1 (0.25) & 0 & 30  & (0.5, 0.5) \\ 
		2  & 5 & 0.1 & 0.8  & 1.2 & 0.5 (0.5) & 0 & 30  & (0.5, 0.5) \\ 
		3  & 5 & 0.1 & 0.8  & 1.2 & 1 (0.25) & 0 & 50  & (0.5, 0.5) \\ 
		4  & 5 & 0.1 & 0.8  & 1.2 & 1 (0.25) & 0 & 20  & (0.5, 0.5) \\ 
		5  & 5 & 0.1 & 0.8  & 1.2 & 1 (0.25) & 0 & 30  & (3, 0.2)   \\ 
		6  & 5 & 0.1 & 0.8  & 1.2 & 1 (0.25) & 0 & 30  & (0.2, 3)   \\ 
		7  & 5 & 0.1 & 0.3  & 0.7 & 1 (0.25) & 0 & 30  & (0.5, 0.5) \\ 
		8  & 5 & 0.1 & 1    & 1   & 1 (0.25) & 0 & 30  & (0.5, 0.5) \\ 
		9  & 5 & 0.5 & 0.8  & 1.2 & 1 (0.25) & 0 & 30  & (0.5, 0.5) \\ 
		10 & 5 & 1   & 0.8  & 1.2 & 1 (0.25) & 0 & 30  & (0.5, 0.5) \\ 
		11 & 1 & 0.1 & 0.8  & 1.2 & 1 (0.25) & 0 & 30  & (0.5, 0.5) \\ 
		12 & 1 & 0.5 & 0.8  & 1.2 & 1 (0.25) & 0 & 30  & (0.5, 0.5) \\ 
		13 & 5 & 0.1 & 0.8  & 1.2 & 1 (0.25) & $2*bw$  & 30  & (0.5, 0.5) \\ 
		14 & 5 & 0.1 & 0.8  & 1.2 & 1 (0.25) & $0.5*bw$ & 30  & (0.5, 0.5) \\ 
		15 & 5 & 0.1 & 0.8  & 1.2 & 1 (0.25) & $2$ & 30 & (0.5, 0.5) \\ 
		16 & 5 & 0.1 & 0.8  & 1.2 & 1 (0.25) & $0.5$ & 30  & (0.5, 0.5) \\  \end{tabular}
	\caption{Overview of runs made in the simulation study for the minimum contrast procedures.  Here $(\sigma^{2}_*, \beta_*)$ denote the start parameters for the optimisation algorithm and $bw$ is the automatically selected bandwidth used to calculate $\hat{g}$ in \texttt{spatstat}.}
	\label{tab:simstudypar}
\end{table}

\subsubsection{Results using the minimum contrast procedure}
In the following we let MCE-$K$ and MCE-$g$ refer to minimum contrast estimation with $T = K$ and $T = g_0$, respectively.  

\begin{figure}
	\centering 
	\includegraphics[width=0.95\textwidth, trim = {0 1cm 0 1cm}, clip]{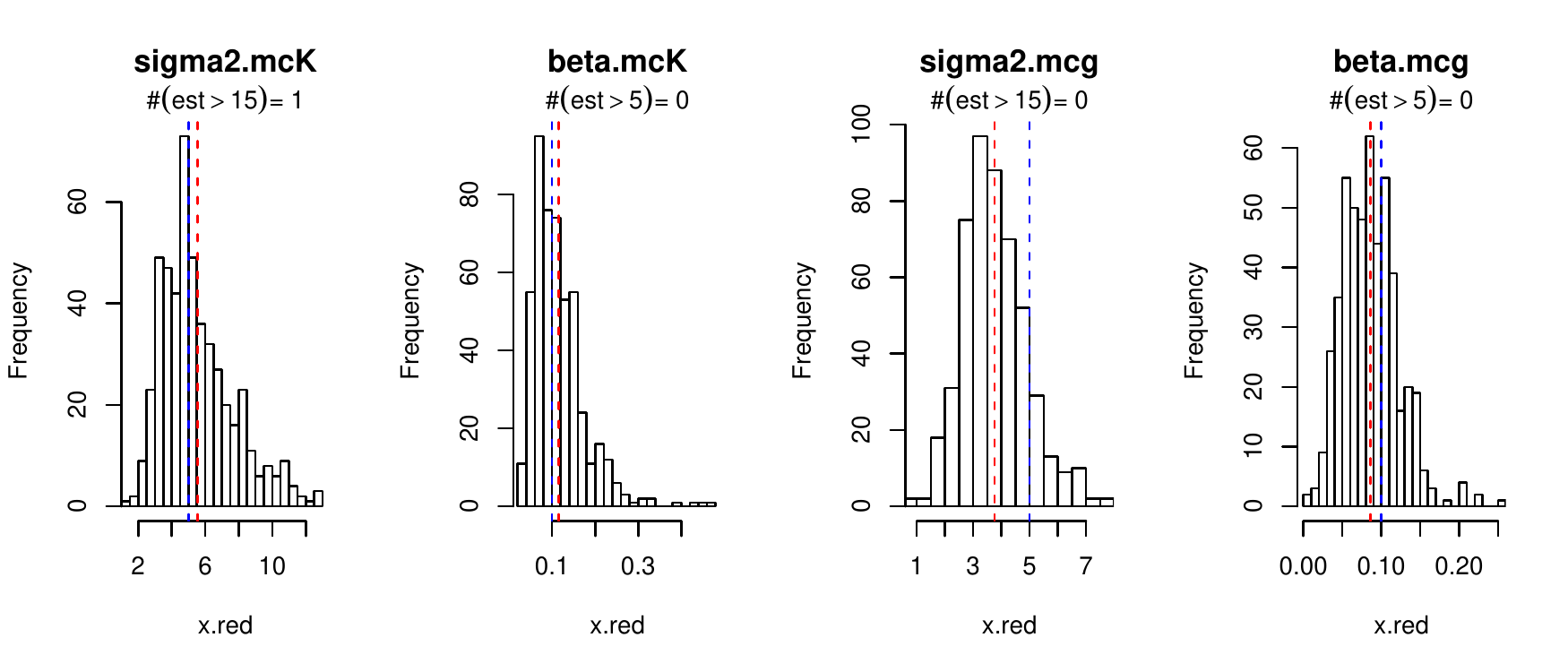}\vspace{4mm}
	\includegraphics[width=0.95\textwidth, trim = {0 1cm 0 1cm}, clip]{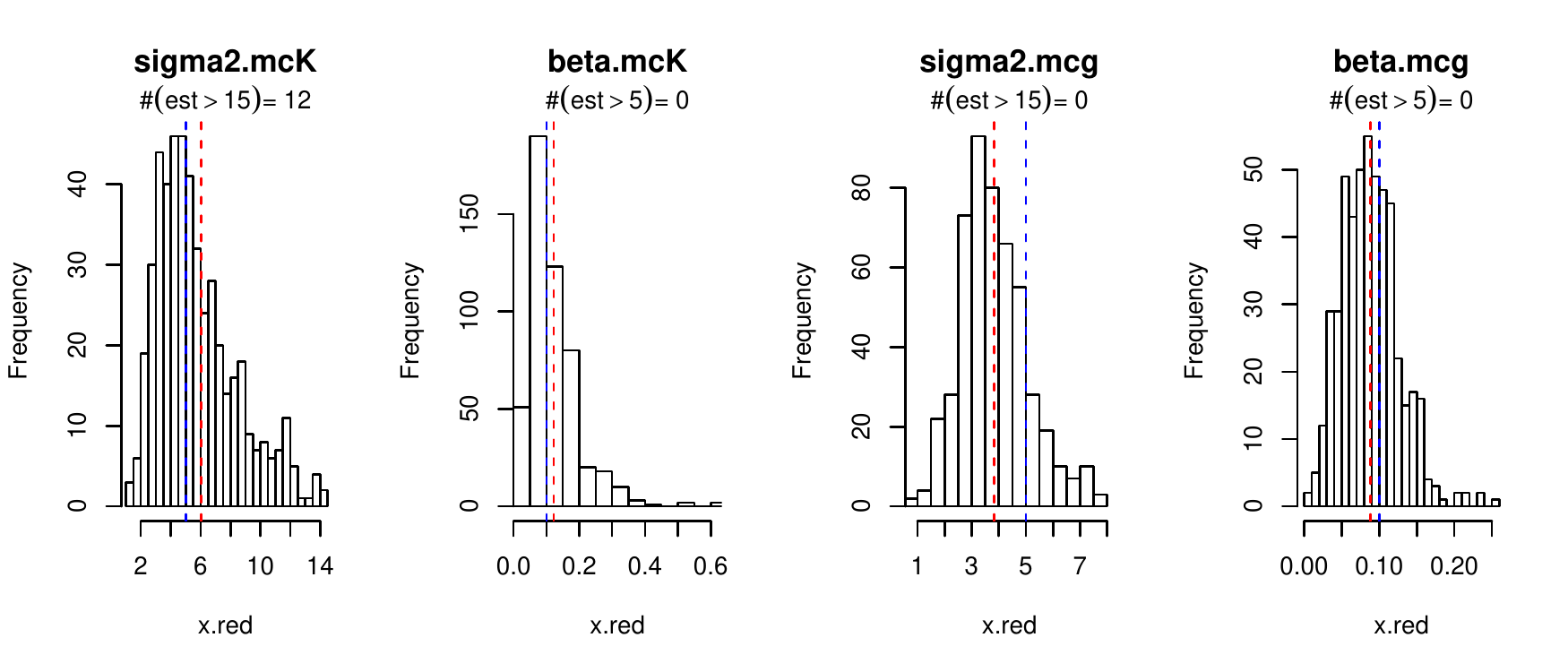}\vspace{4mm}
	\includegraphics[width=0.95\textwidth, trim = {0 1cm 0 1cm}, clip]{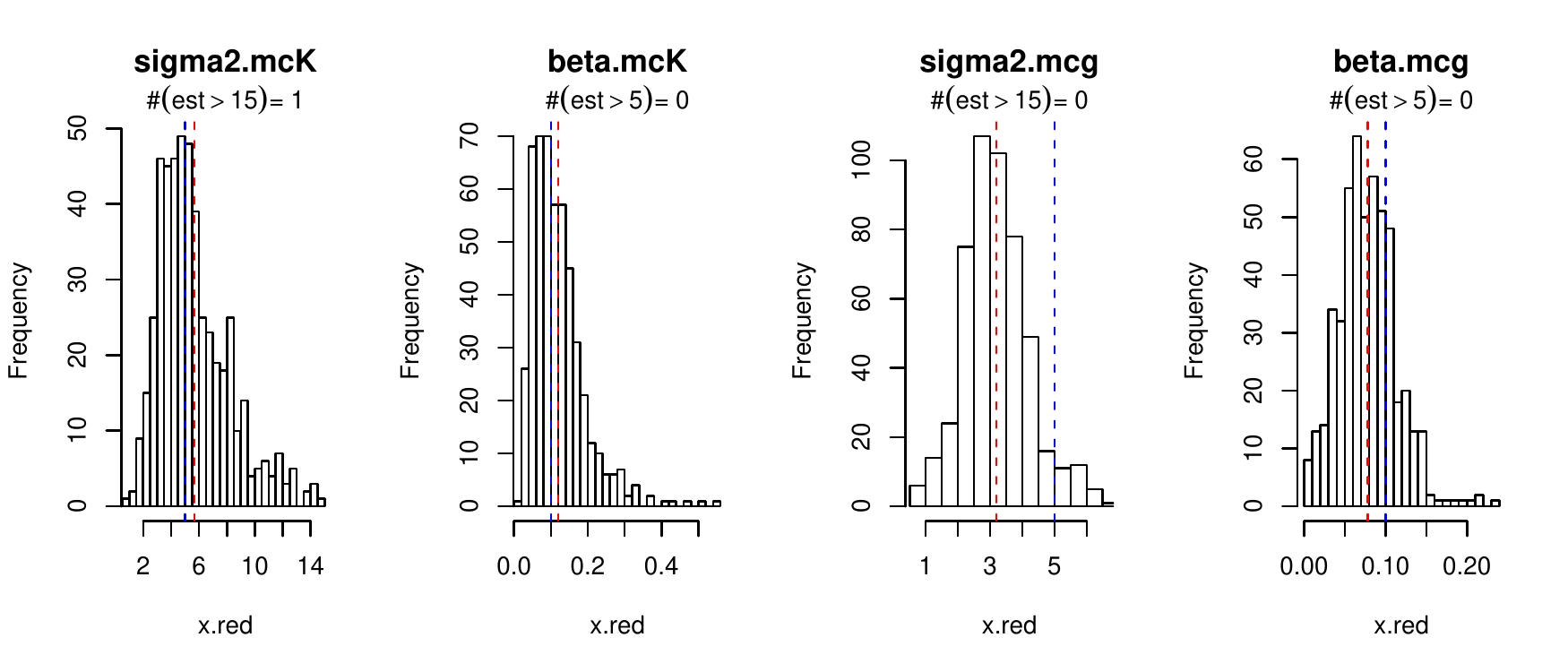}\vspace{4mm}
	\includegraphics[width=0.95\textwidth, trim = {0 1cm 0 1cm}, clip]{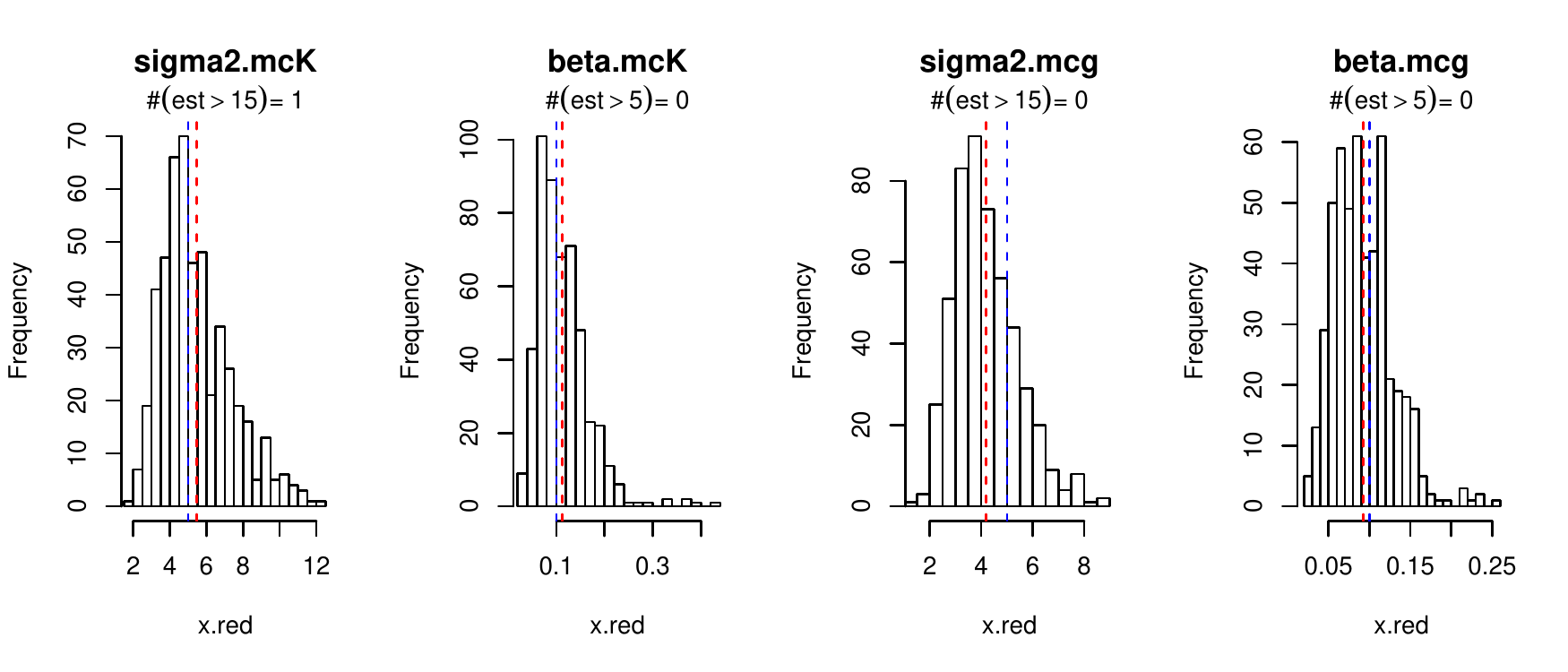}
	\caption{Estimates of $\sigma^2$ and $\beta$ using either MCE-$g$ or MCE-$K$ for 500 simulated point patterns of models with parameters no.\ 1--4 in Table~\ref{tab:simstudypar} (one set of parameters for each row, starting with no.\ 1 in the top). From left to right: estimates of $\sigma^2$ and $\beta$ found by MCE-$K$ (column 1 and 2), followed by estimates of $\sigma^2$ and $\beta$ based on MCE-$g$ (column 3 and 4). Blue dashed line is the true parameter value, and red dashed line is the mean of the estimates. OBS: the histograms have been truncated such that estimates above 15 for $\sigma^2$ (column 1 and 3) and 5 for $\beta$ (column 2 and 4) have been omitted in the frequency count; in each histogram it is stated how many values were discarded.}
	\label{fig:simstudy1_4} 
\end{figure}

\begin{figure}
	\centering 
	\includegraphics[width=0.95\textwidth, trim = {0 1cm 0 1cm}, clip]{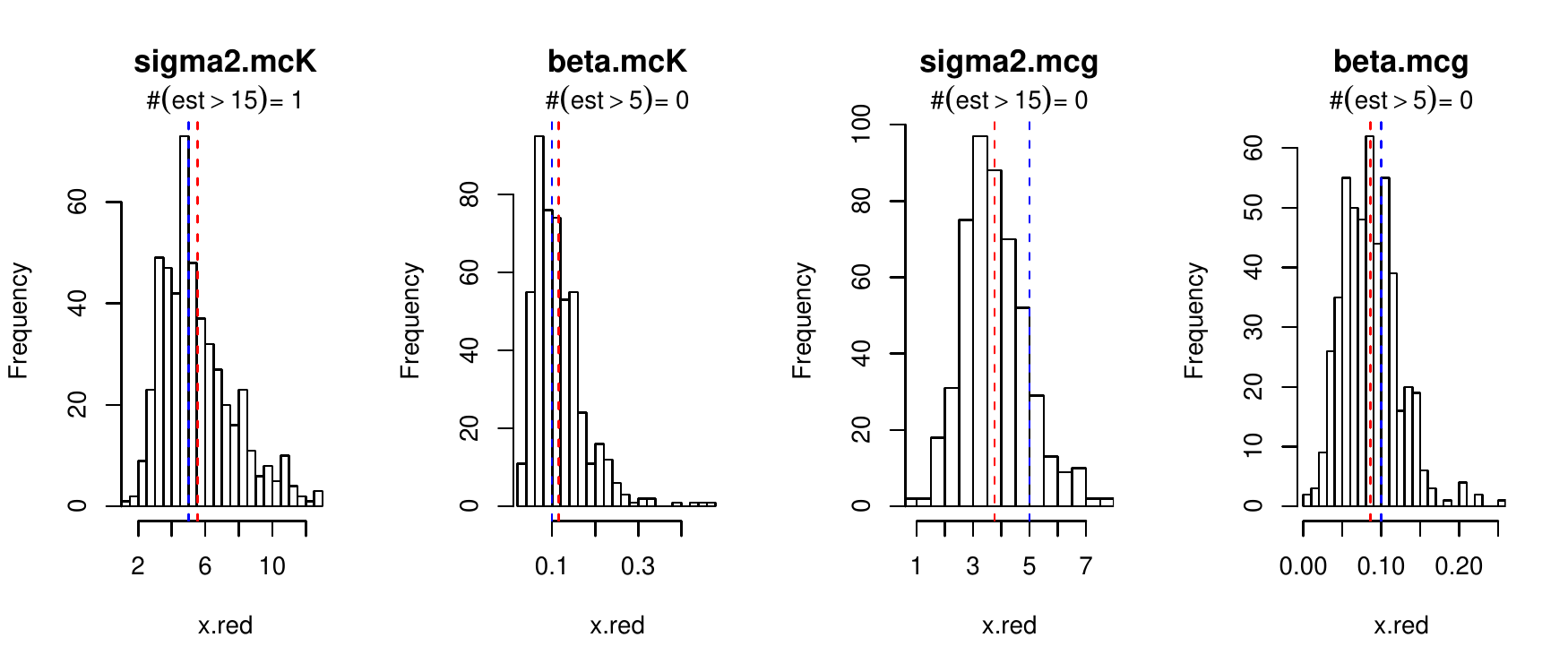}\vspace{4mm}
	\includegraphics[width=0.95\textwidth, trim = {0 1cm 0 1cm}, clip]{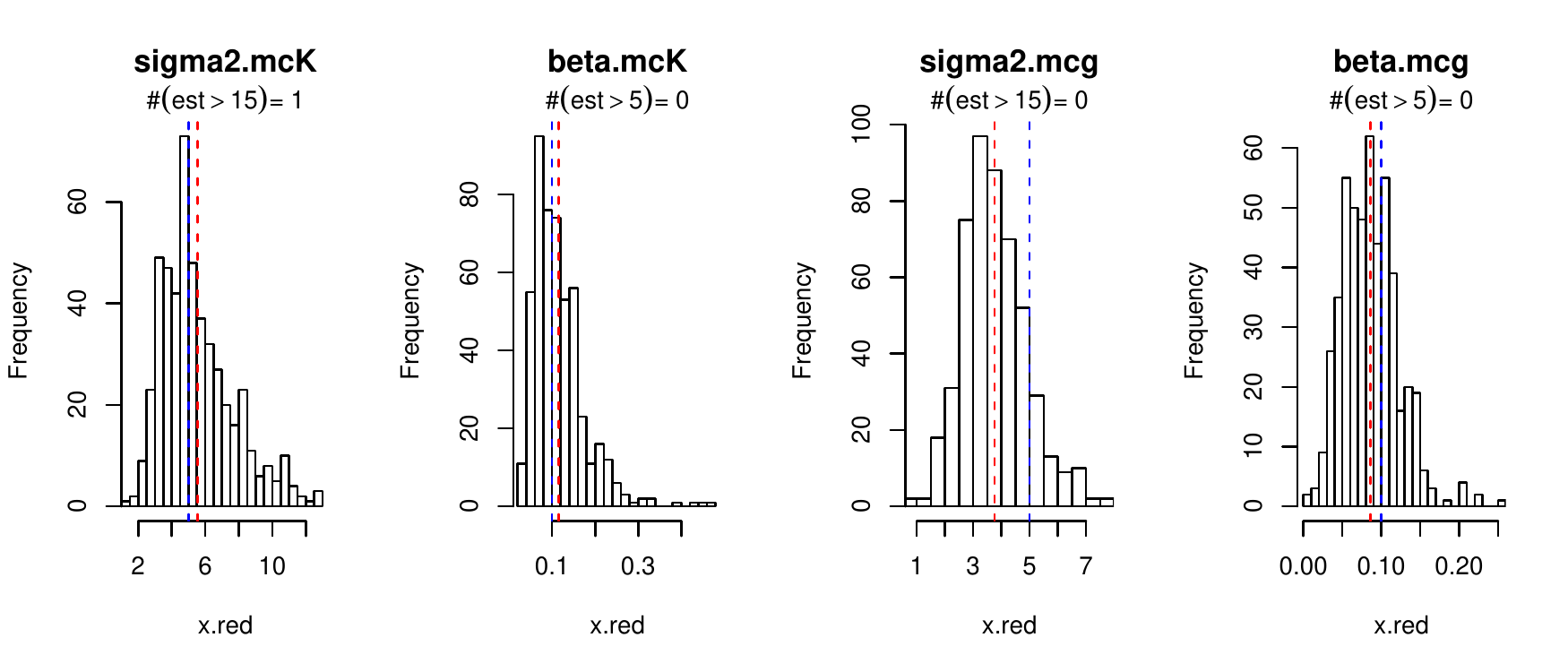}\vspace{4mm}
	\includegraphics[width=0.95\textwidth, trim = {0 1cm 0 1cm}, clip]{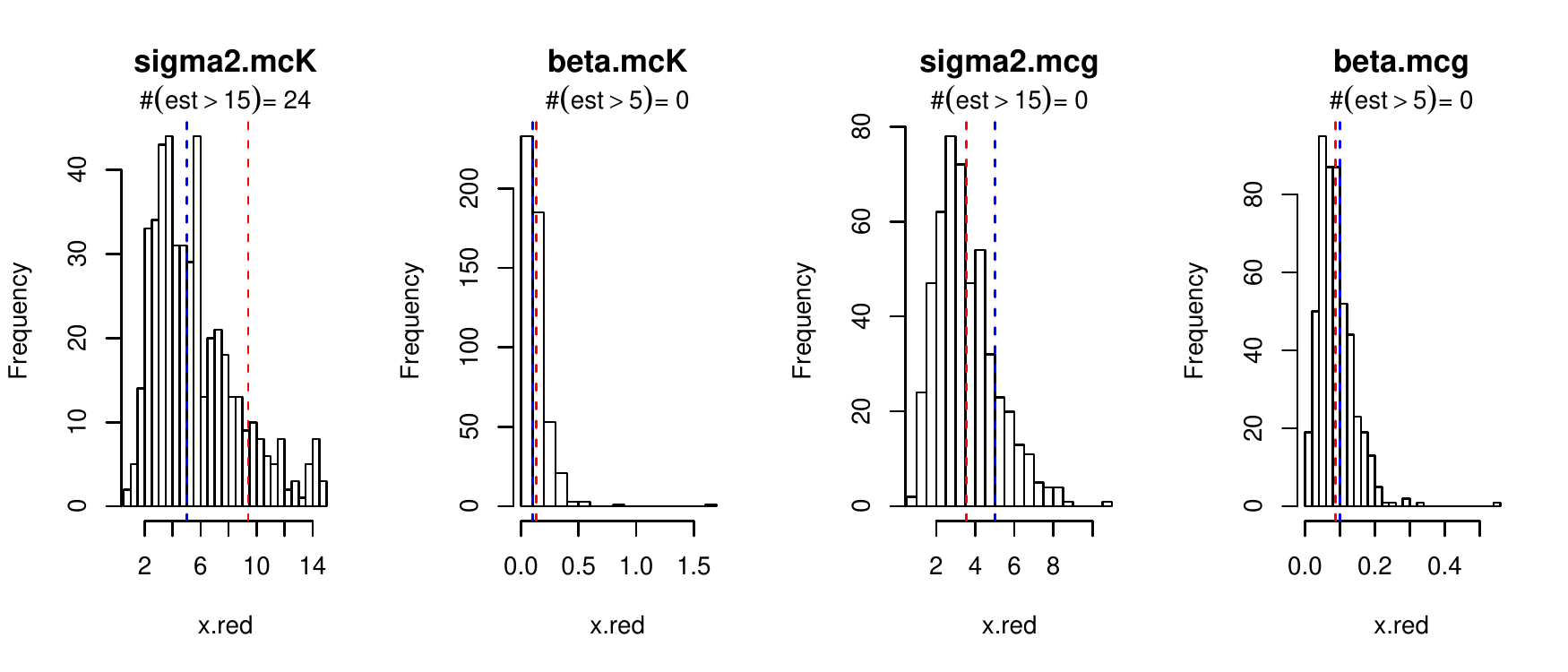}\vspace{4mm}
	\includegraphics[width=0.95\textwidth, trim = {0 1cm 0 1cm}, clip]{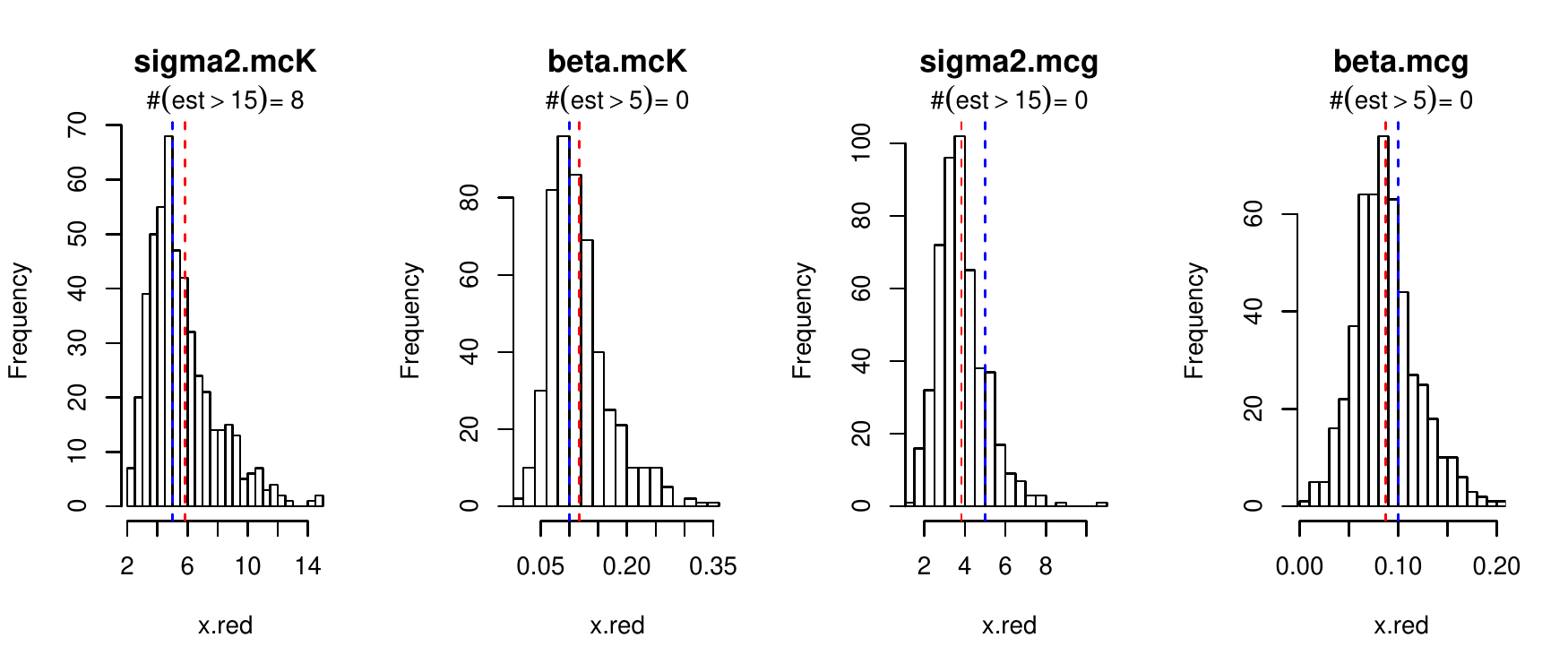}
	\caption{Estimates of $\sigma^2$ and $\beta$ using either MCE-$g$ or MCE-$K$ for 500 simulated point patterns of models with parameters no.\ 5--8 in Table~\ref{tab:simstudypar} (one set of parameters for each row, starting with no.\ 5 in the top). From left to right: estimates of $\sigma^2$ and $\beta$ found by MCE-$K$ (column 1 and 2), followed by estimates of $\sigma^2$ and $\beta$ based on MCE-$g$ (column 3 and 4). Blue dashed line is the true parameter value, and red dashed line is the mean of the estimates. OBS: the histograms have been truncated such that estimates above 15 for $\sigma^2$ (column 1 and 3) and 5 for $\beta$ (column 2 and 4) have been omitted in the frequency count; in each histogram it is stated how many values were discarded.}
	\label{fig:simstudy5_8}
\end{figure}

\begin{figure}
	\centering 
	\includegraphics[width=0.95\textwidth, trim = {0 1cm 0 1cm}, clip]{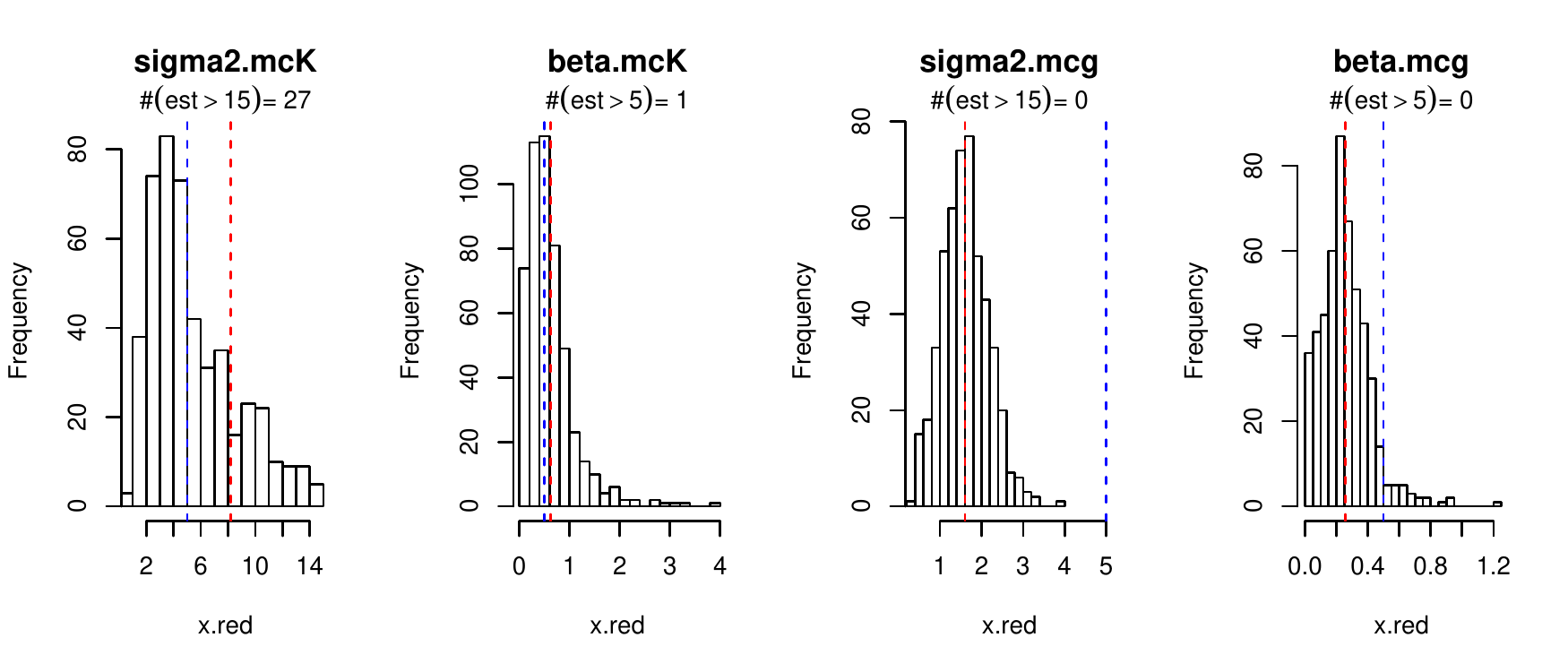}\vspace{4mm}
	\includegraphics[width=0.95\textwidth, trim = {0 1cm 0 1cm}, clip]{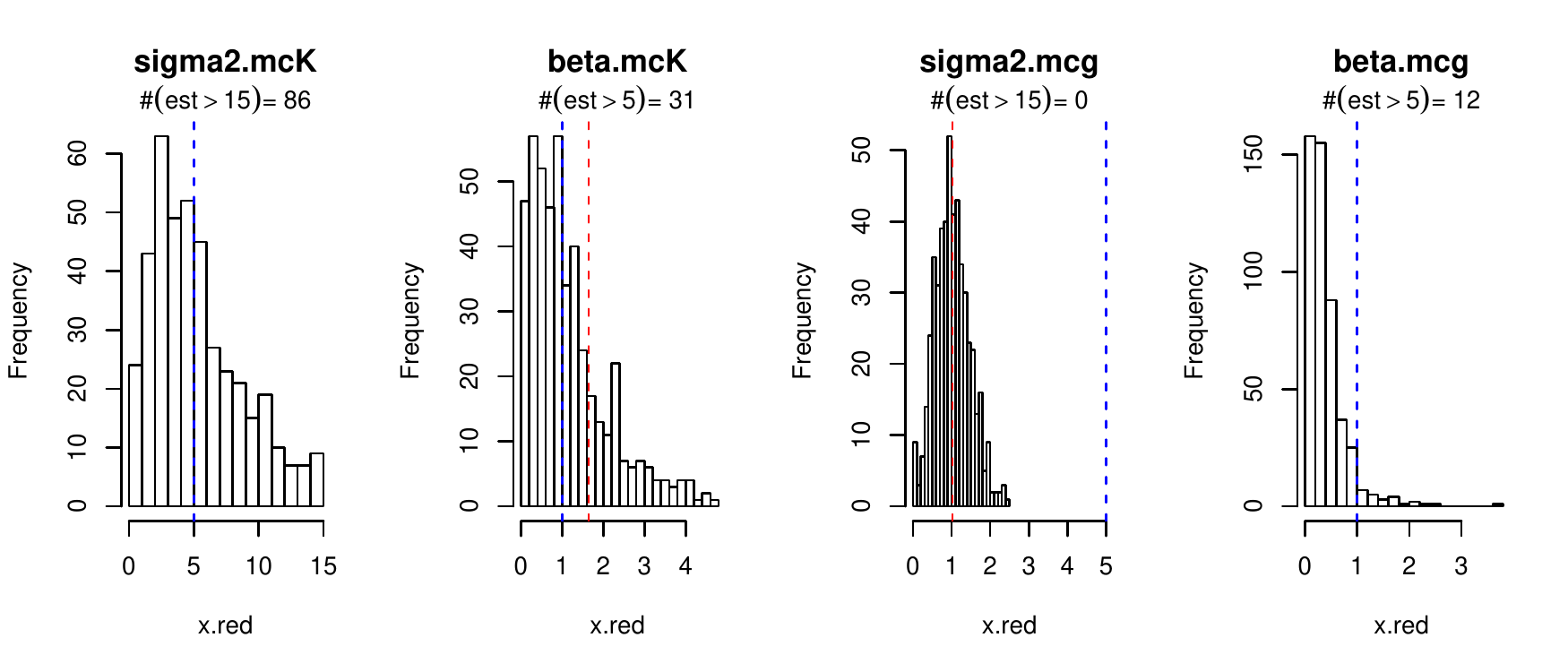}\vspace{4mm}
	\includegraphics[width=0.95\textwidth, trim = {0 1cm 0 1cm}, clip]{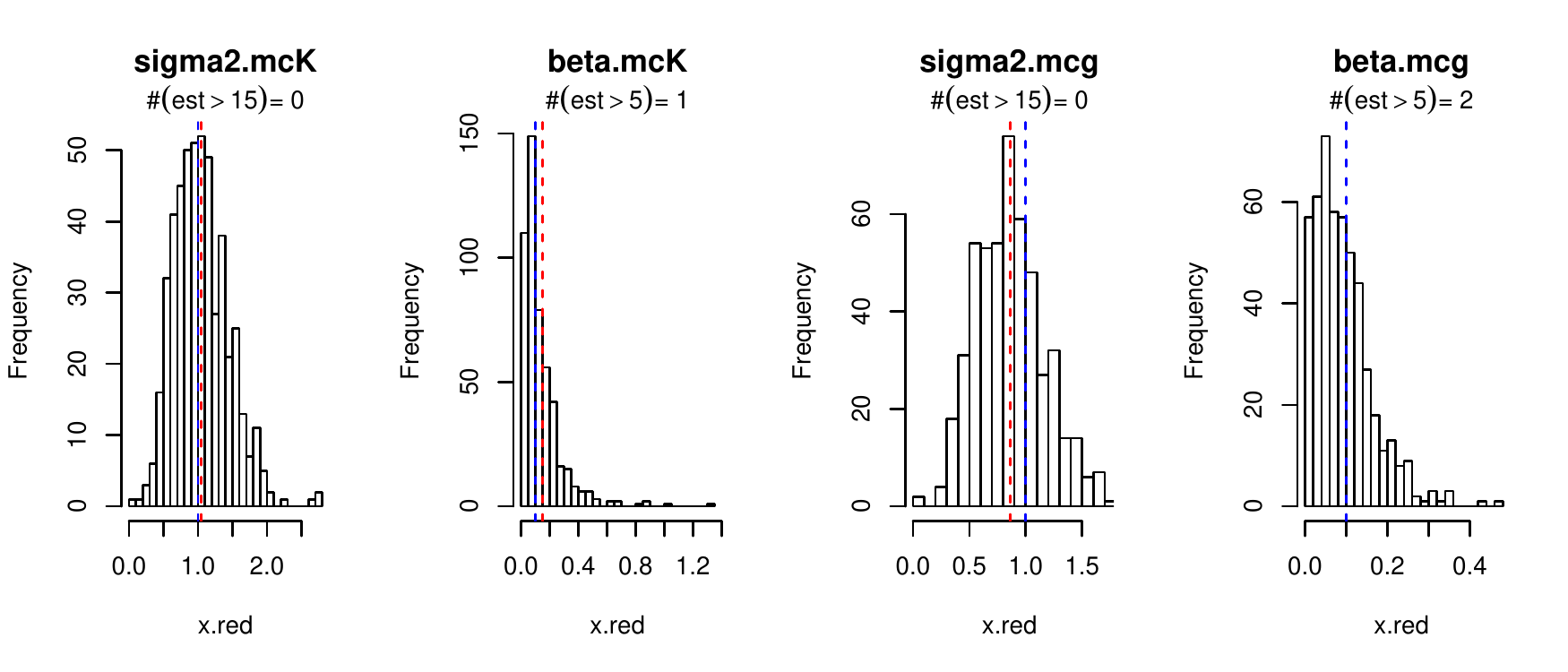}\vspace{4mm}
	\includegraphics[width=0.95\textwidth, trim = {0 1cm 0 1cm}, clip]{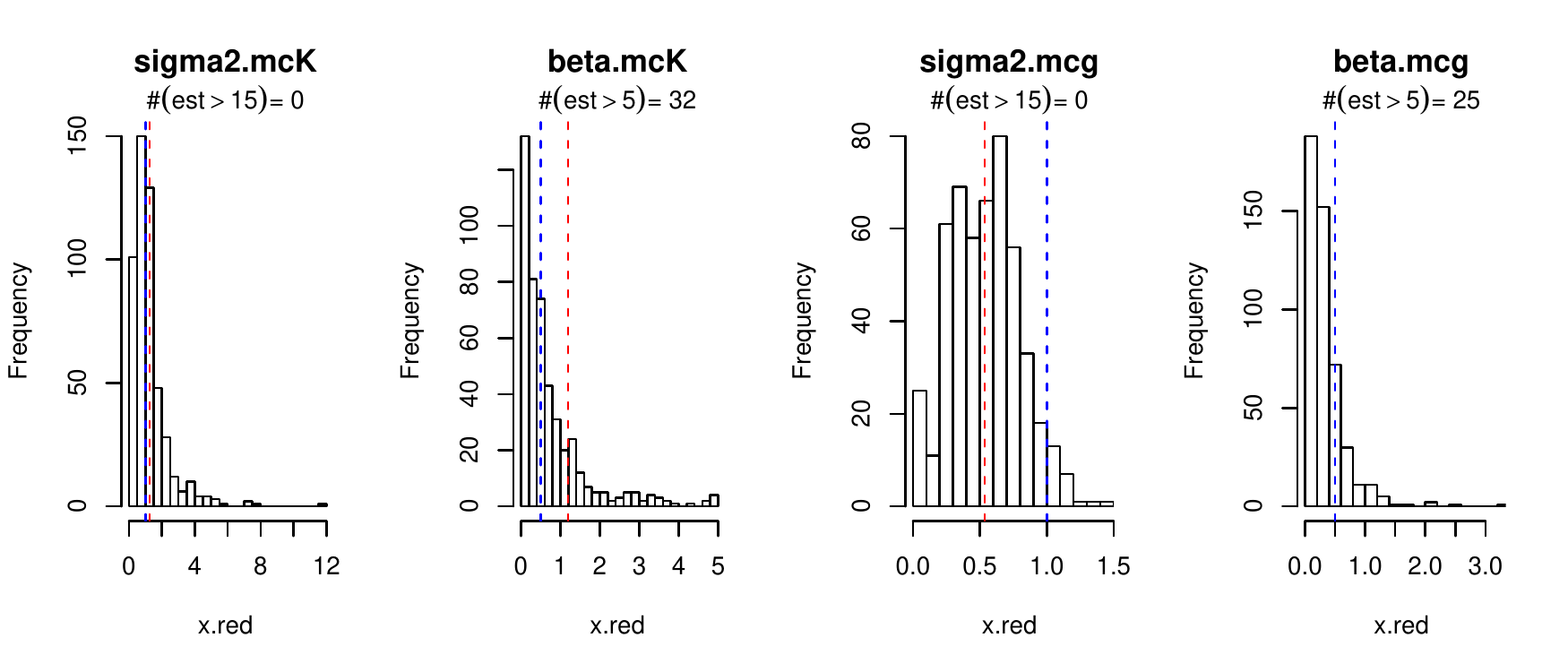}
	\caption{Estimates of $\sigma^2$ and $\beta$ using either MCE-$g$ or MCE-$K$ for 500 simulated point patterns of models with parameters no.\ 9--12 in Table~\ref{tab:simstudypar} (one set of parameters for each row, starting with no.\ 9 in the top). From left to right: estimates of $\sigma^2$ and $\beta$ found by MCE-$K$ (column 1 and 2), followed by estimates of $\sigma^2$ and $\beta$ based on MCE-$g$ (column 3 and 4). Blue dashed line is the true parameter value, and red dashed line is the mean of the estimates. OBS: the histograms have been truncated such that estimates above 15 for $\sigma^2$ (column 1 and 3) and 5 for $\beta$ (column 2 and 4) have been omitted in the frequency count; in each histogram it is stated how many values were discarded.}
	\label{fig:simstudy9_12}
\end{figure}

\begin{figure}
	\centering 
	\includegraphics[width=0.95\textwidth, trim = {0 1cm 0 1cm}, clip]{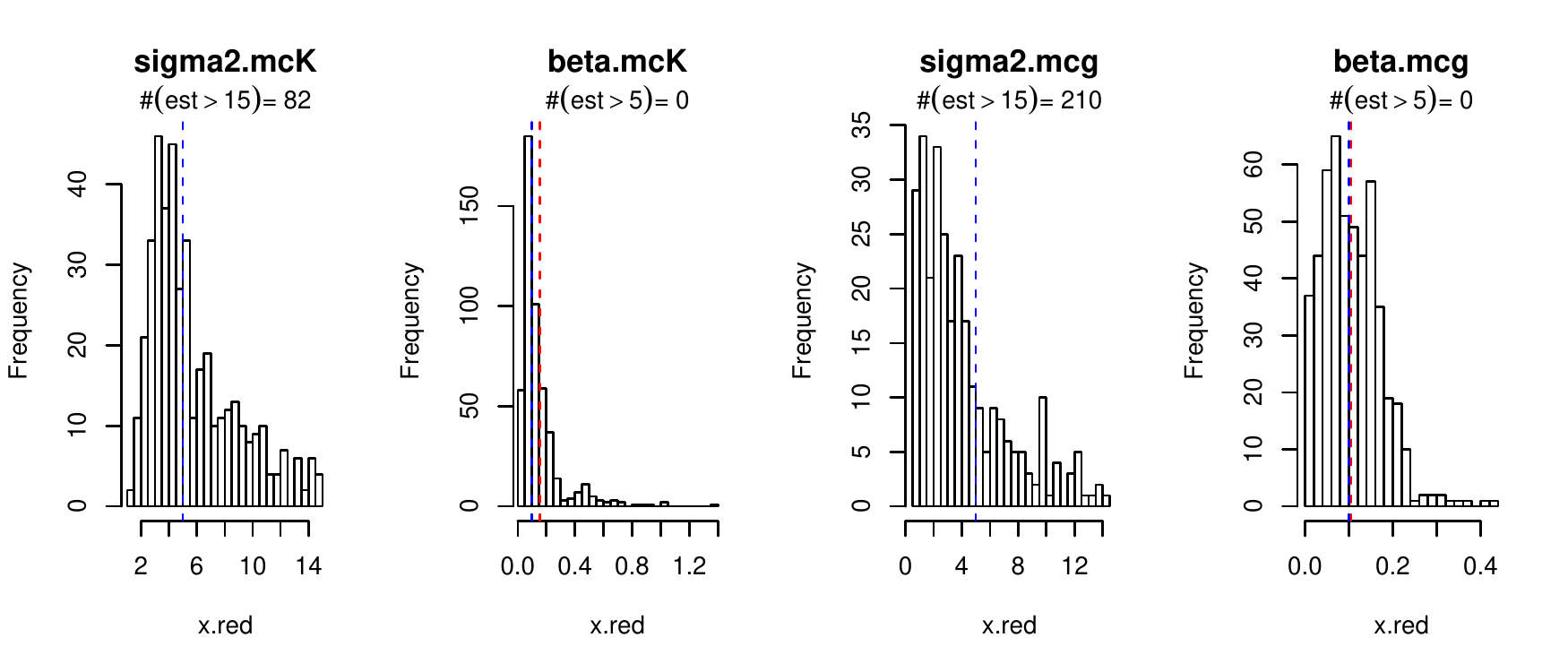}\vspace{4mm}
	\includegraphics[width=0.95\textwidth, trim = {0 1cm 0 1cm}, clip]{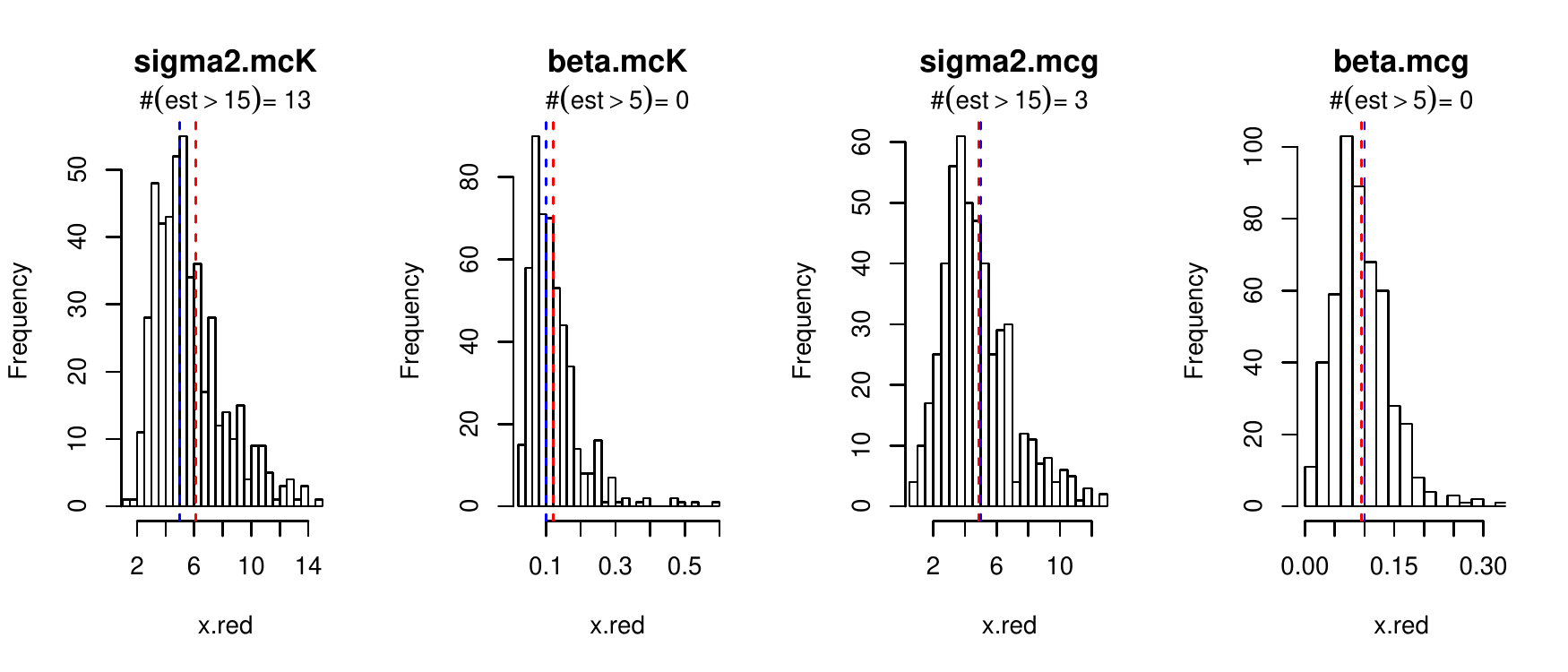}\vspace{4mm}
	\includegraphics[width=0.95\textwidth, trim = {0 1cm 0 1cm}, clip]{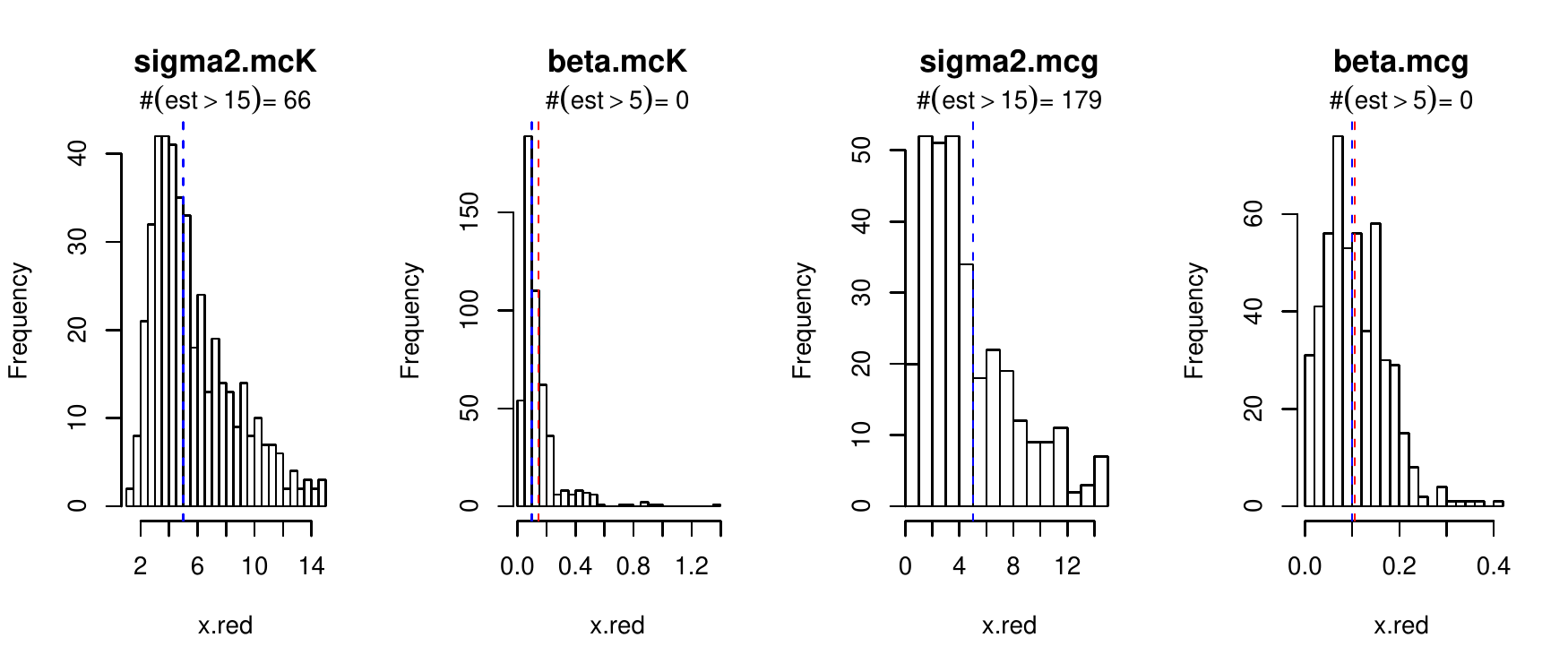}\vspace{4mm}
	\includegraphics[width=0.95\textwidth, trim = {0 1cm 0 1cm}, clip]{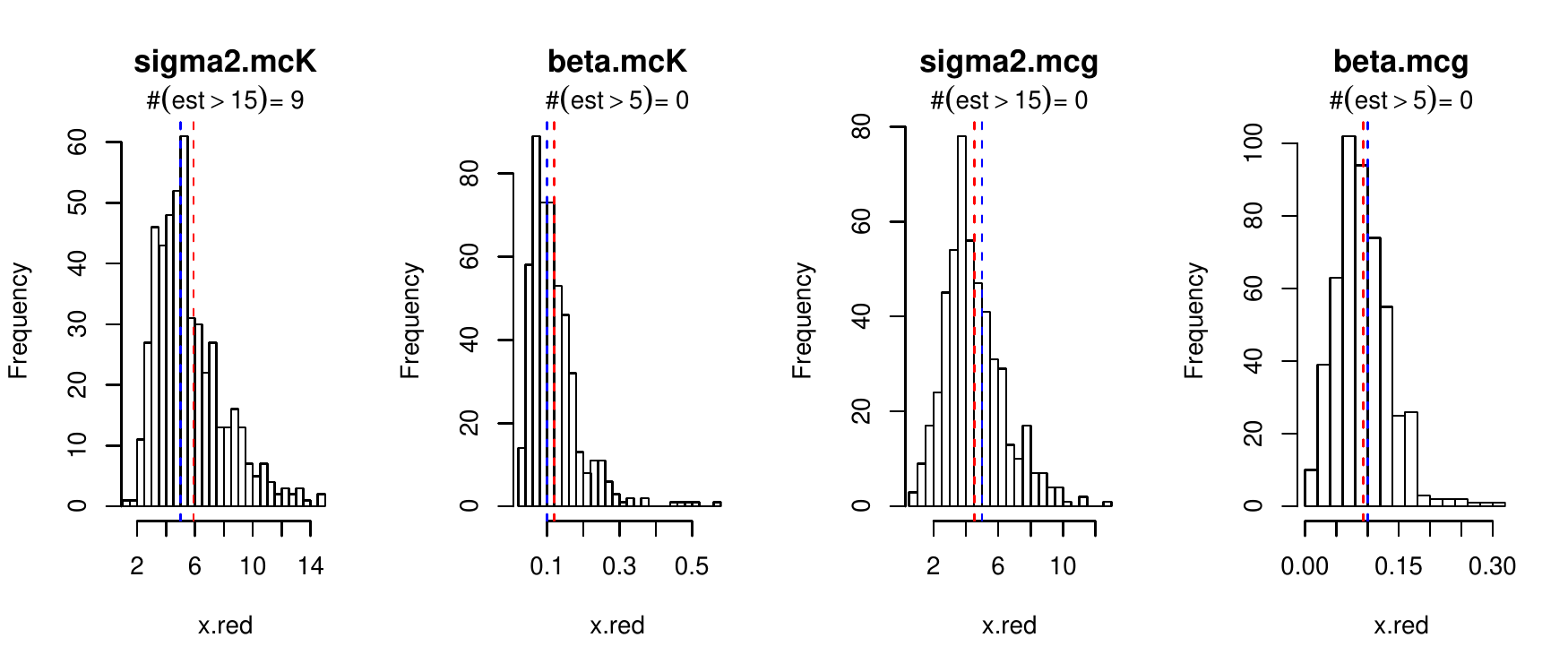}
	\caption{Estimates of $\sigma^2$ and $\beta$ using either MCE-$g$ or MCE-$K$ for 500 simulated point patterns of models with parameters no.\ 13--16 in  Table~\ref{tab:simstudypar} (one set of parameters for each row, starting with no.\ 13 in the top). From left to right: estimates of $\sigma^2$ and $\beta$ found by MCE-$K$ (column 1 and 2), followed by estimates of $\sigma^2$ and $\beta$ based on MCE-$g$ (column 3 and 4). Blue dashed line is the true parameter value, and red dashed line is the mean of the estimates. OBS: the histograms have been truncated such that estimates above 15 for $\sigma^2$ (column 1 and 3) and 5 for $\beta$ (column 2 and 4) have been omitted in the frequency count
		; in each histogram it is stated how many values were discarded.}
	\label{fig:simstudy13_16}
\end{figure}

Histograms of the obtained estimates are shown in Figures~\ref{fig:simstudy1_4}--\ref{fig:simstudy13_16}.
In general it seems that MCE-$g$ performs better than MCE-$K$. For parameter values that result in models close to the Poisson model, neither of the estimation procedures estimate $(\sigma^2, \beta)$ successfully. 
For simulations more distinguishable from Poisson (as for example run no.\ 1), 
the MCE-$g$ gives more satisfactory estimates. 

Neither MCE-$g$ or MCE-$K$ seem to be sensitive to the choice of start parameters for the reference run. Further, for the reference run it seem that $r_u = 20$ was the best choice, but in general this depend on the model we are trying to fit and naturally on the size of the network. 
The best choice of $r_l$ seem to be $r_l = 0$, while $p = 0.25$ seems preferable over $p = 0.5$ for MCE-$K$, and $p = 0.5$ and $p = 1$ perform equally well for the MCE-$g$. 
The choice of $r_u$ seem to be important with respect to bias and variance: a too high $r_u$ may lead to a large bias, while a too smale $r_u$ may lead to a large variance in the estimates. It is therefore recommendable to perform a small simulation study for the specific network and proposed model at hand, such that the best choice of $r_u$ can be made. 

\subsubsection{Results using the adaptive composite likelihood}
\begin{figure}
	\centering 
	\includegraphics[width=0.95\textwidth, trim = {0 1cm 0 1cm}, clip]{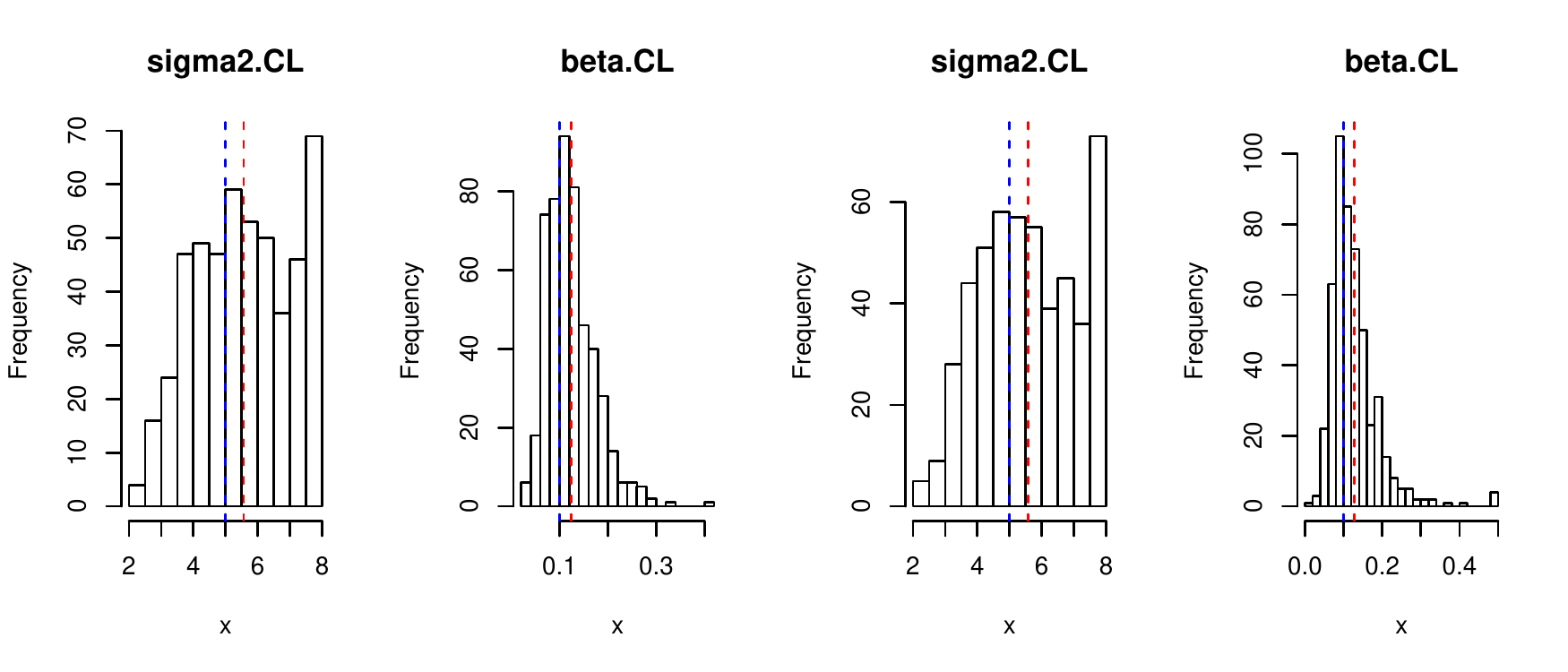}\vspace{5mm}
	\includegraphics[width=0.95\textwidth, trim = {0 1cm 0 1cm}, clip]{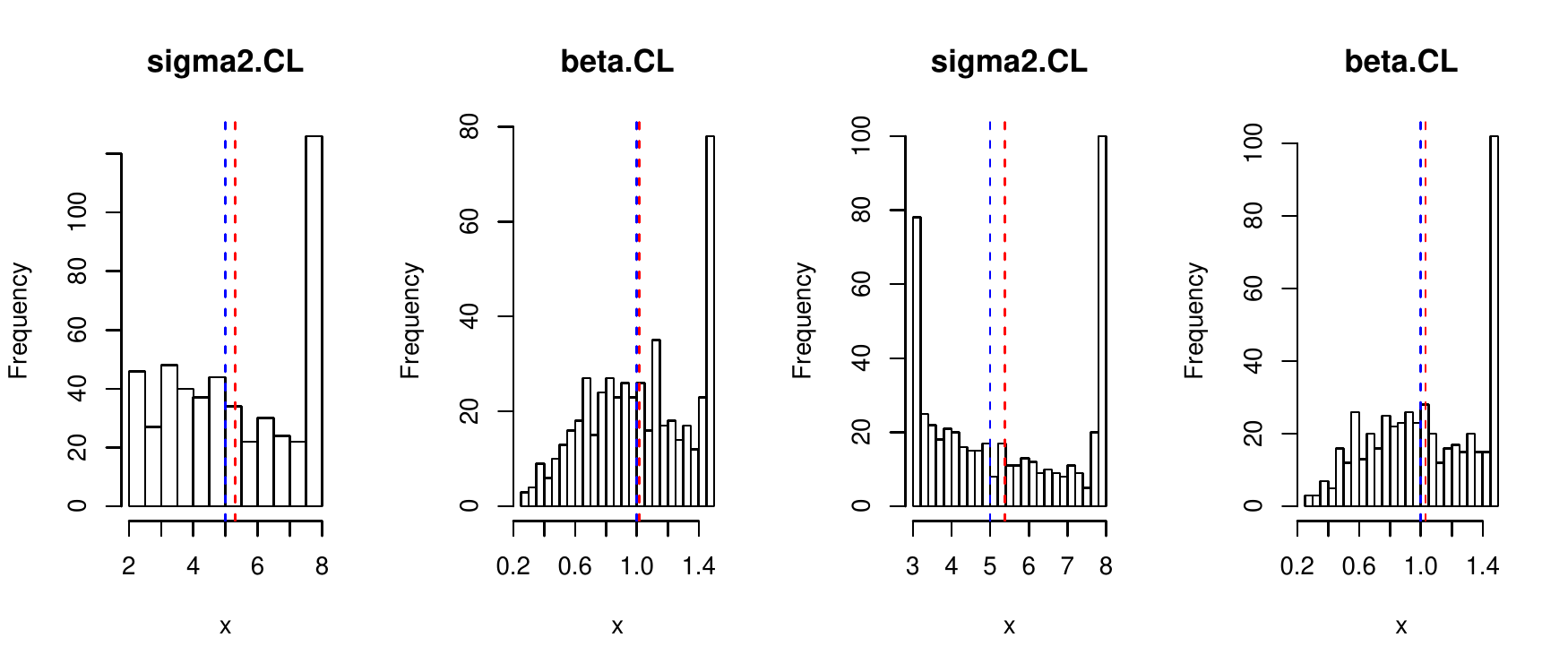}\vspace{5mm}
	\includegraphics[width=0.95\textwidth, trim = {0 1cm 0 1cm}, clip]{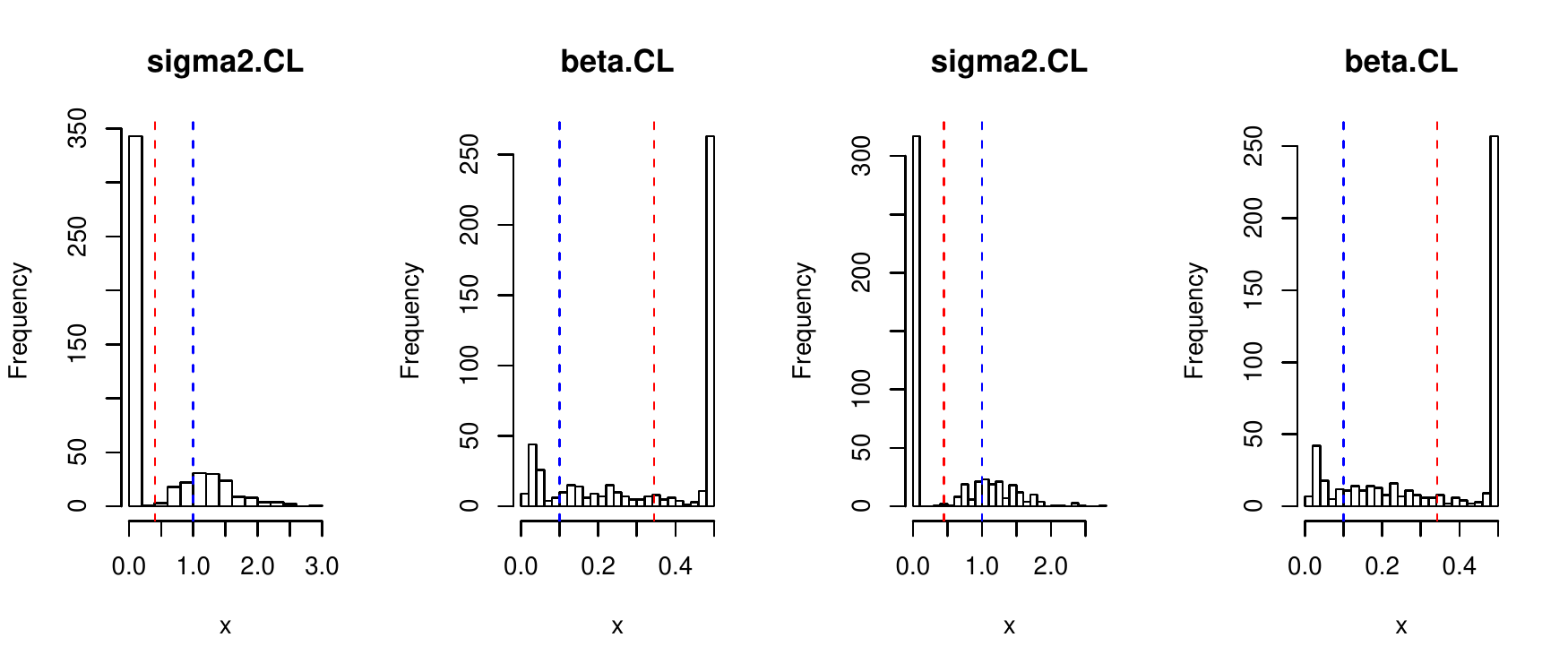}\vspace{5mm}
	\includegraphics[width=0.95\textwidth, trim = {0 1cm 0 1cm}, clip]{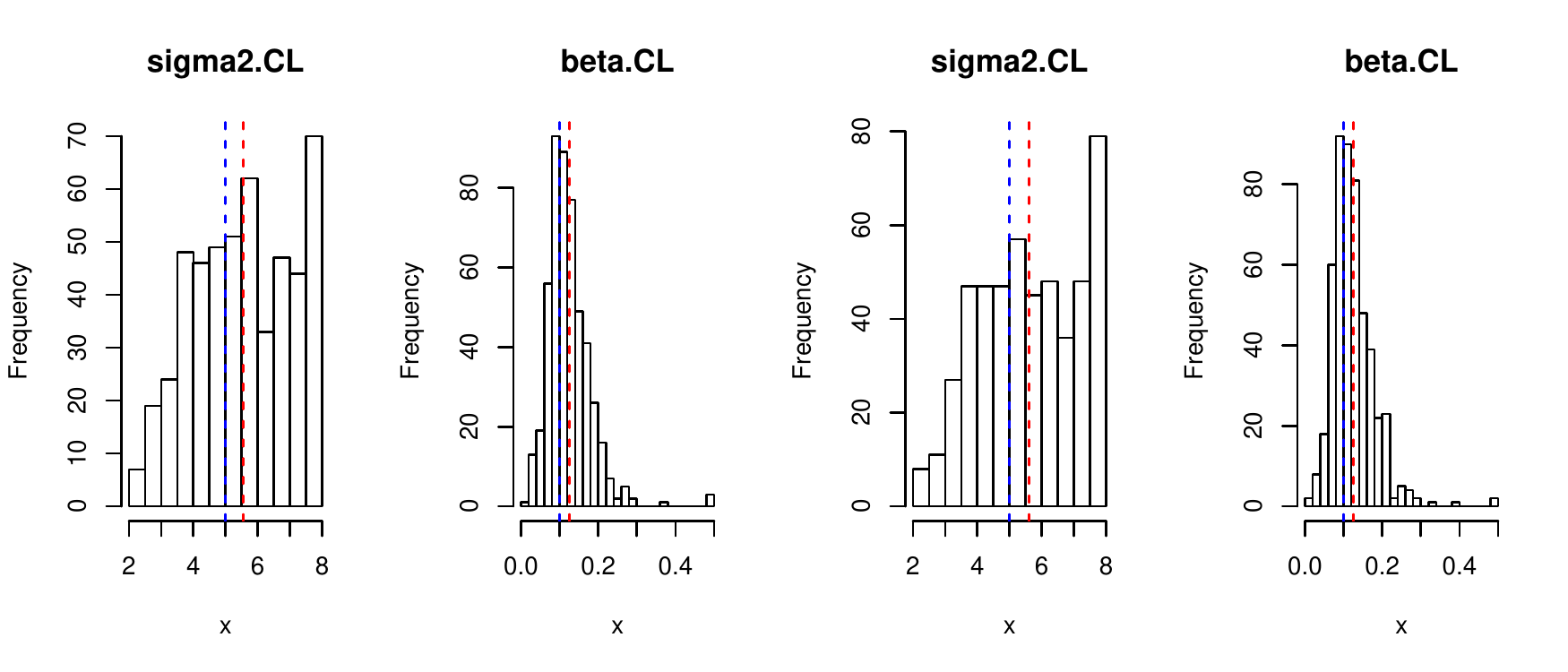}
	\caption{Results from simulation study using adaptive composite likelihood: estimates of $\beta$ and $\sigma^2$ for model parameters from run no.\ 1 (first row), 10 (second row), and 11 (third row) with $\epsilon = 0.05$ and run no.\ 1 with $\epsilon = 0.01$ (fourth row); see Table~\ref{tab:simstudypar}. The two first columns display estimates of $\sigma^2$ and $\beta$ (in that order) when using the indicator weight function for the CLE procedure, while estimates in the two right columns are found using CLE with the exponential weight function. Blue dashed line is the true parameter value, red dashed line is the mean of the estimates.}
	\label{fig:simstudyCLE1} 
\end{figure}

For the CLE procedure we restricted ourselves to a small number of runs as the grid search was very time consuming. Specifically, we simulated 500 point patterns from the Cox process models with the parameters from run no.\ 1, 10, and 11 in Table~\ref{tab:simstudypar}. For each choice of model parameters we estimated $(\sigma^2, \beta)$ using both weight functions and $\epsilon = 0.05$. Further, we also ran the CLE procedure with $\epsilon = 0.01$ for the model parameters from run no.\ 1.  Figure~\ref{fig:simstudyCLE1} shows histograms of the resulting estimates. It is clear that for all three choices of model parameters the grid should be broader in order to find the parameter values that give the smallest length of \eqref{eq:score_RW} and that the estimates are worse than the ones obtained by the minimum contrast procedures. Finally, there is no seemingly advantage of choosing one weight function over the other or of choosing $\epsilon = 0.01$ over $\epsilon = 0.05$. 

\section{Analysis of spine locations}\label{app:C}
\begin{figure}
\centering
\includegraphics[width=\textwidth]{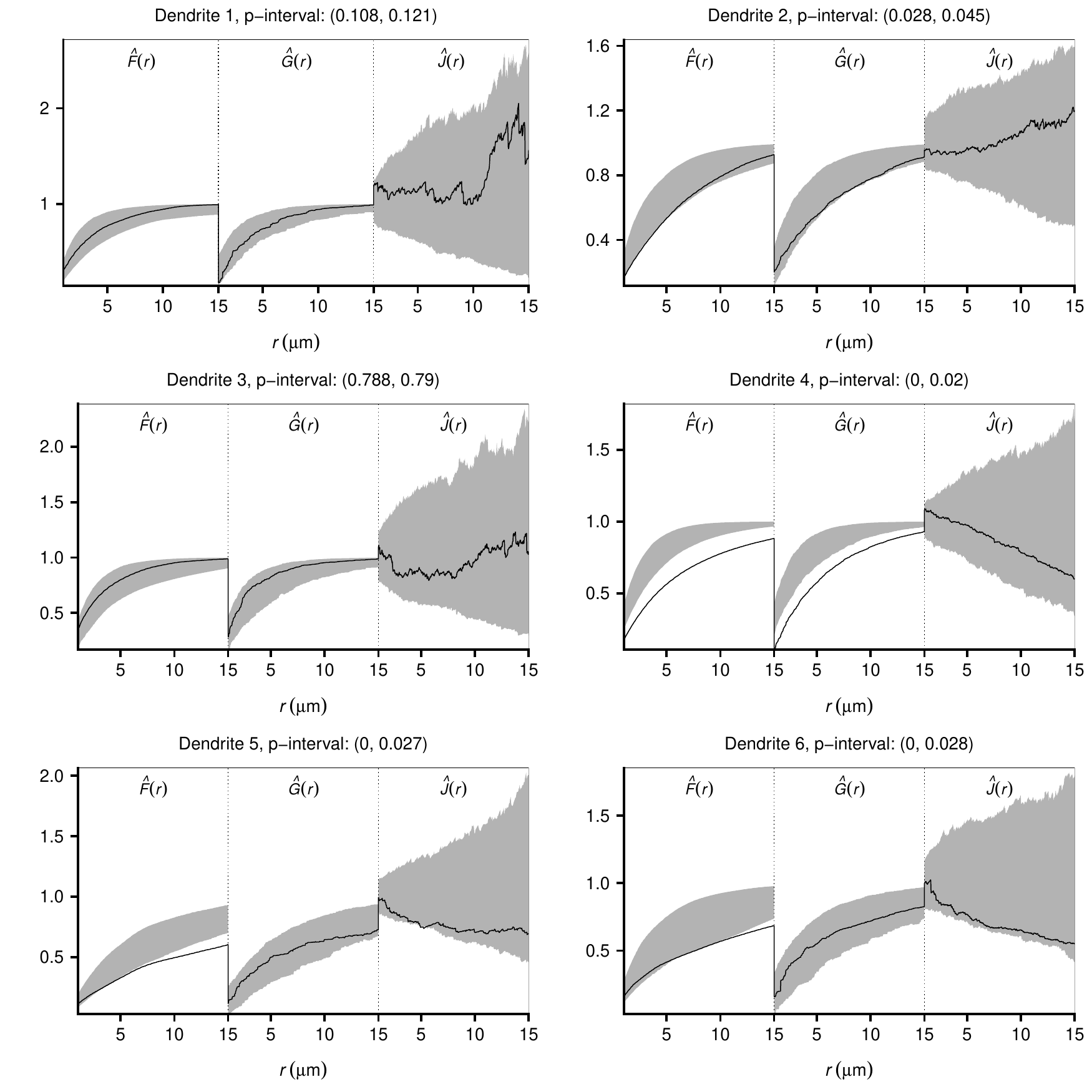}
\caption{For each spine data set: the concatenation of $\hat{F}$, $\hat{G}$, and $\hat{J}$ for the spine locations (black solid line) along with $95\%$ global rank envelopes (grey region) based on 2499 simulations from the fitted inhomogeneous Poisson model; $p$-intervals for each of the associated global rank envelope tests are also displayed. Here $r$-values less than $\SI{1}{\micro \meter}$ are disregarded.}
\label{fig:JFGPoisson}
\end{figure}

\begin{figure}
	\centering
	\includegraphics[width=0.9\textwidth]{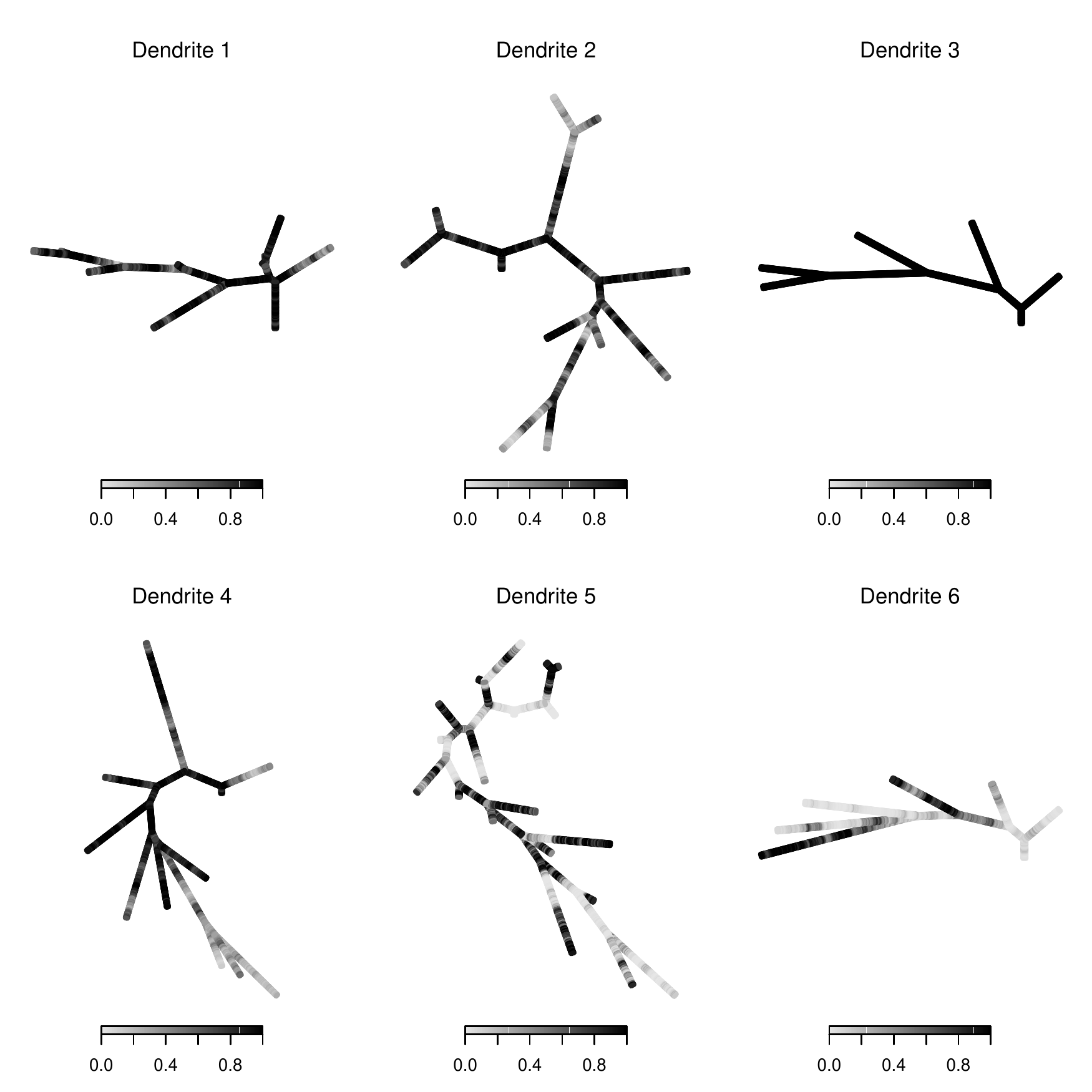}
	\caption{For each dendrite tree, a simulated realisation of the random field $\Pi$ determining the retention probabilities in the fitted Cox process models.}
	\label{fig:RF_simulation}
\end{figure}

\begin{figure}
	\centering
	\includegraphics[width=\textwidth]{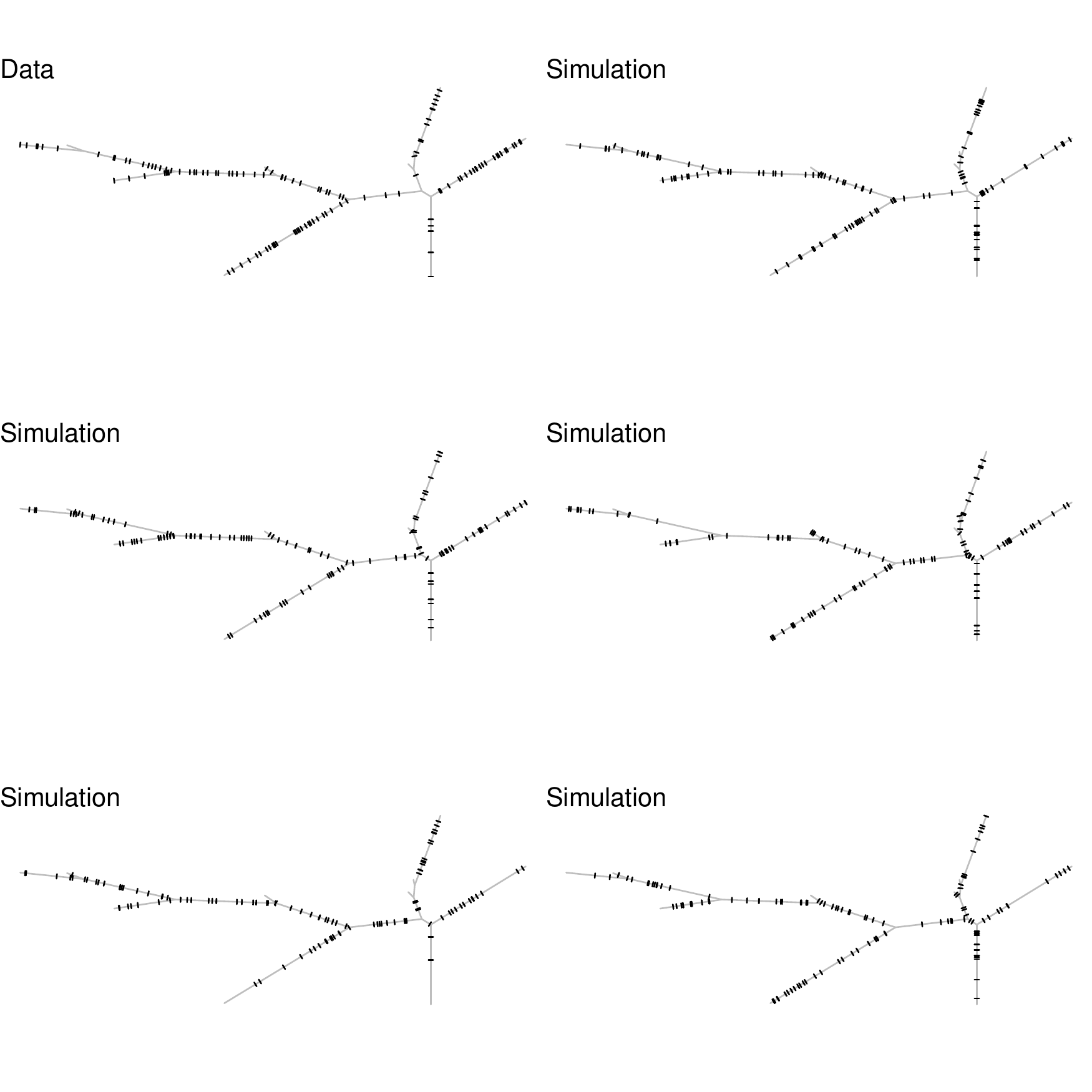}
	\caption{Upper left corner: observed spine locations on dendrite 1. Remaining: simulated point patterns from the fitted Cox process model. The simulated retention probabilities used to obtain the point pattern in the upper right corner are shown in  Figure~\ref{fig:RF_simulation}.} 
	\label{fig:neuron1_simulations}
\end{figure}

\begin{figure}
	\centering
	\includegraphics[width=\textwidth]{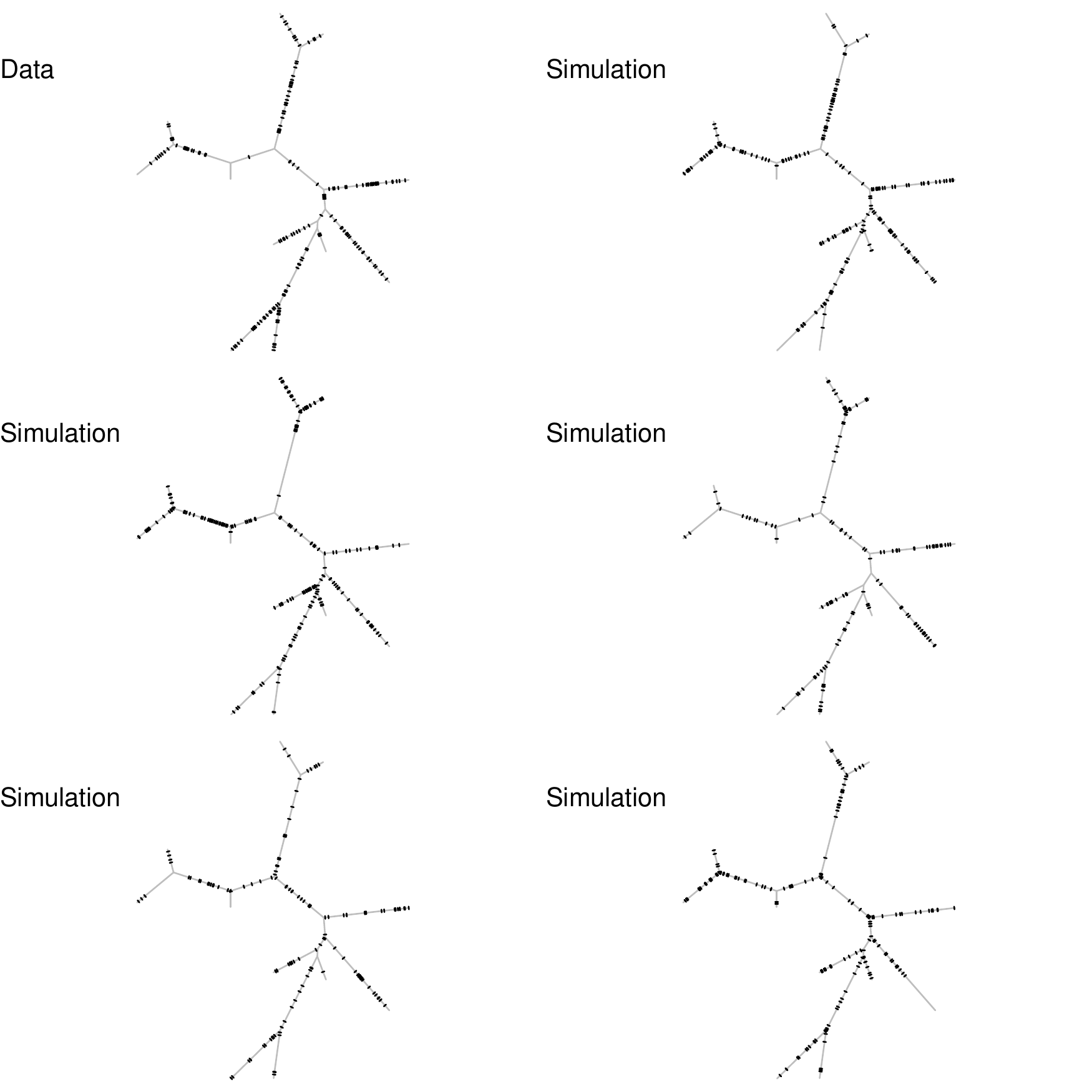}
	\caption{Upper left corner: observed spine locations on dendrite 2. Remaining: simulated point patterns from fitted Cox process model. The simulated retention probabilities used to obtain the point pattern in the upper right corner are shown in  Figure~\ref{fig:RF_simulation}.}
	\label{fig:neuron2_simulations}
\end{figure}

\begin{figure}
	\centering
	\includegraphics[width=\textwidth]{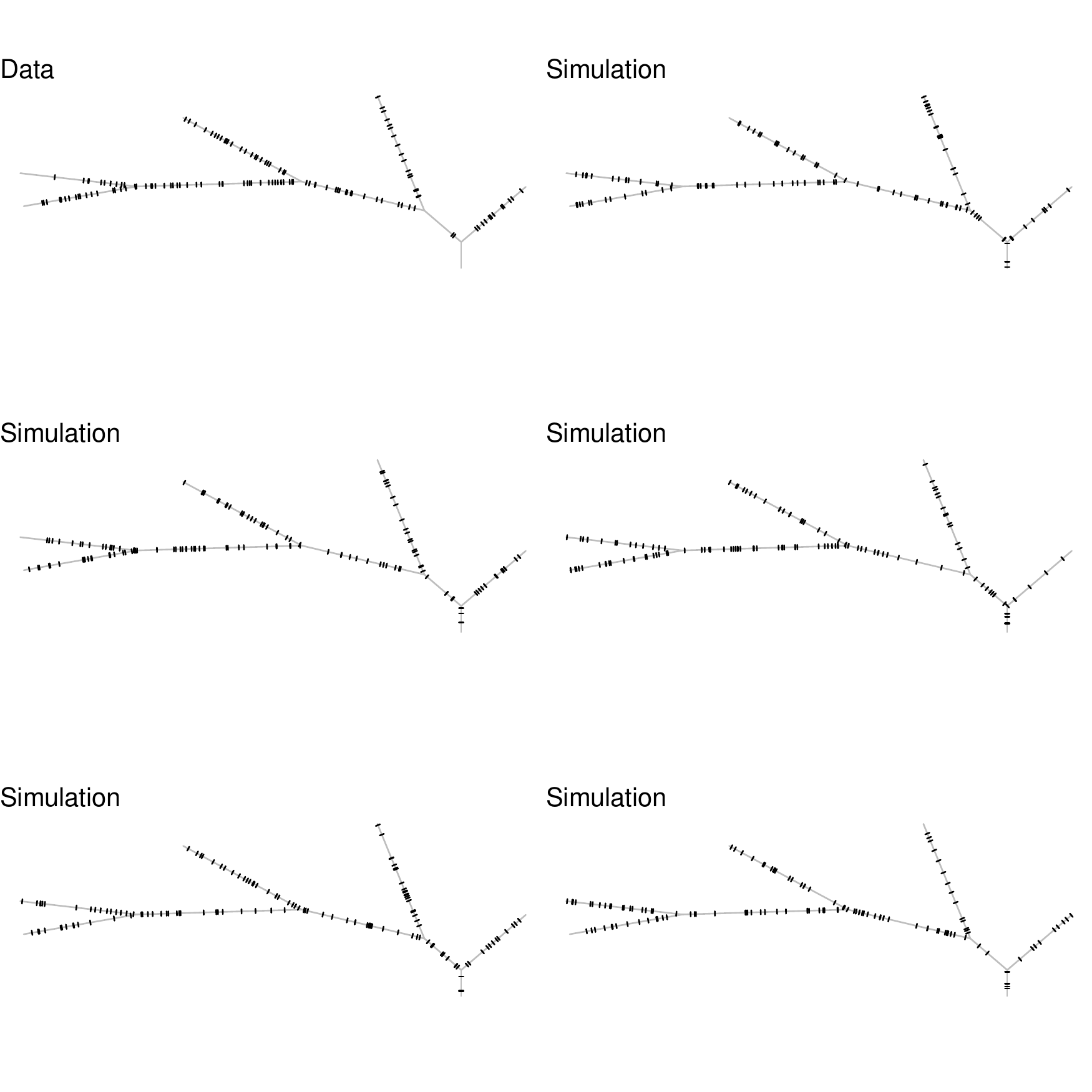}
	\caption{Upper left corner: observed spine locations on dendrite 3. Remaining: simulated point patterns from fitted Cox process model. The simulated retention probabilities used to obtain the point pattern in the upper right corner are shown in Figure~\ref{fig:RF_simulation}.}
	\label{fig:neuron3_simulations}
\end{figure}

\begin{figure}
	\centering
	\includegraphics[width=\textwidth]{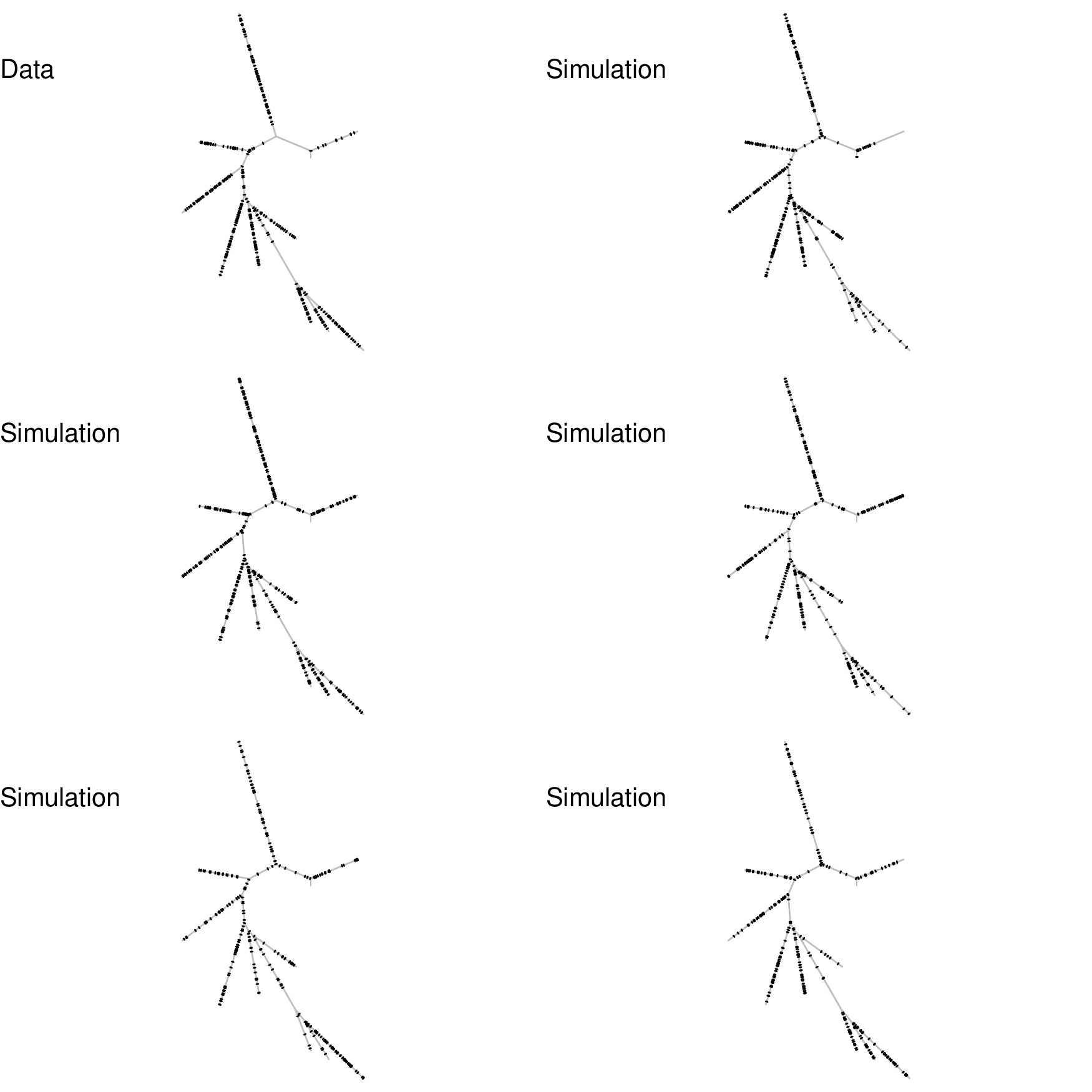}
	\caption{Upper left corner: observed spine locations on dendrite 4. Remaining: simulated point patterns from fitted Cox process model. The simulated retention probabilities used to obtain the point pattern in the upper right corner are shown in Figure~\ref{fig:RF_simulation}.}
	\label{fig:neuron4_simulations}
\end{figure}

\begin{figure}
	\centering
	\includegraphics[width=\textwidth]{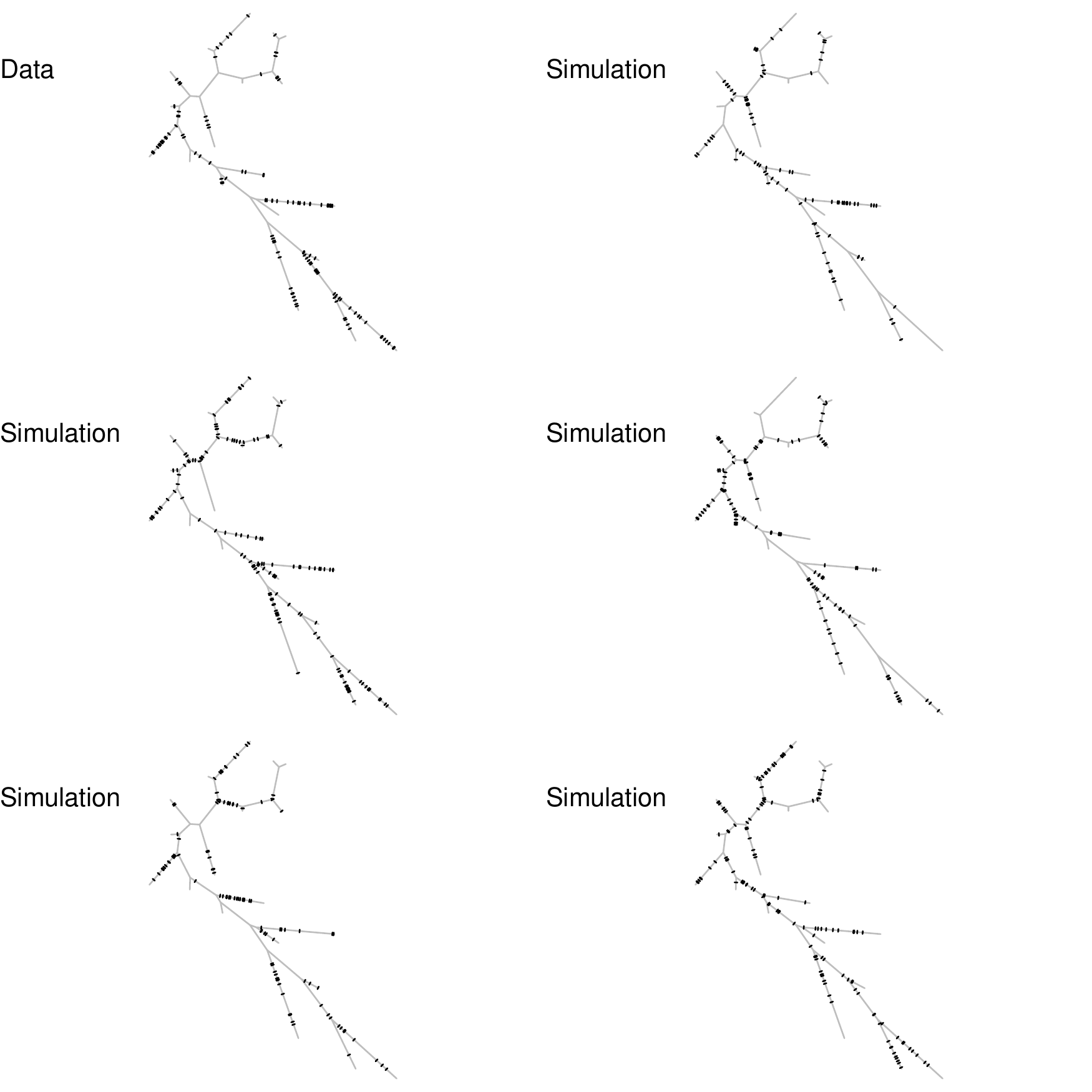}
	\caption{Upper left corner: observed spine locations on dendrite 5. Remaining: simulated point patterns from fitted Cox process model. The simulated retention probabilities used to obtain the point pattern in the upper right corner are shown in Figure~\ref{fig:RF_simulation}.}
	\label{fig:neuron5_simulations}
\end{figure}

\begin{figure}
	\centering
	\includegraphics[width=\textwidth]{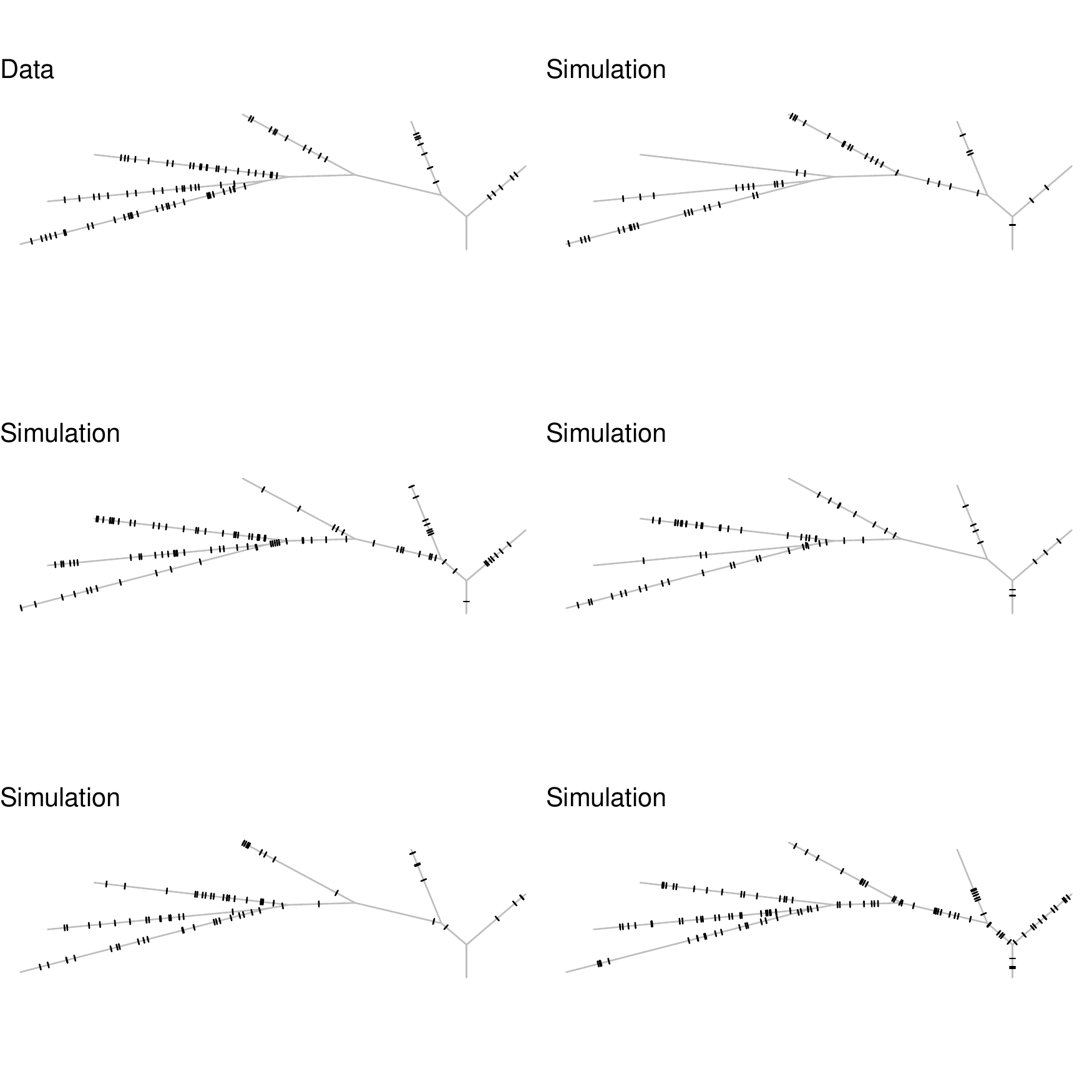}
	\caption{Upper left corner: observed spine locations on dendrite 6. Remaining: simulated point patterns from fitted Cox process model. The simulated retention probabilities used to obtain the point pattern in the upper right corner are shown in  Figure~\ref{fig:RF_simulation}.}
	\label{fig:neuron6_simulations}
\end{figure}

\begin{figure}
	\centering
	\includegraphics[width=\textwidth]{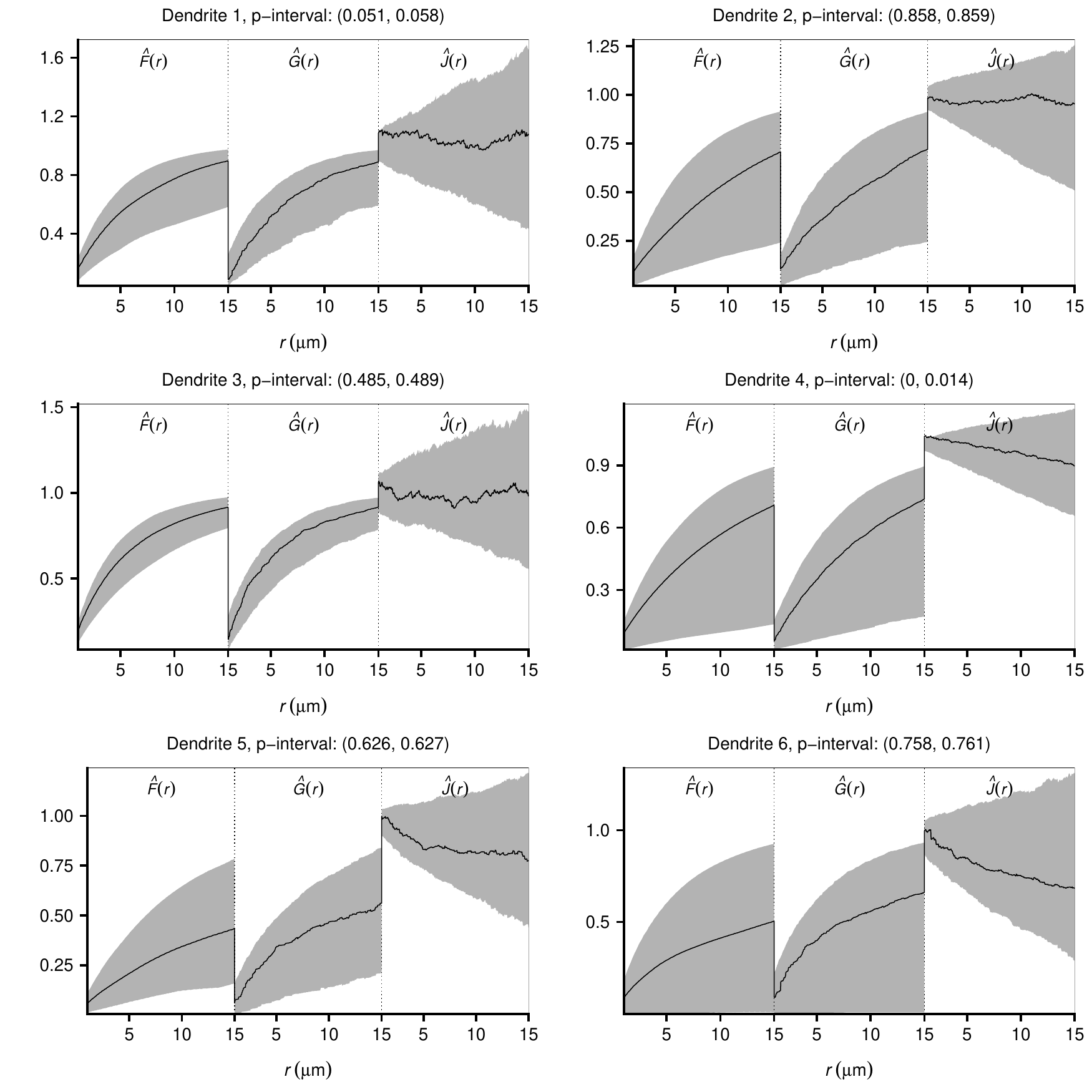}
	\caption{For each spine data set: the concatenation of the estimated $F$-, $G$-, and $J$-functions from \cite{cronie-etal-19} for the spine locations (black solid line) along with $95\%$ global rank envelopes (grey region) based on 2499 simulations from the fitted Cox model; $p$-intervals for each of the associated global rank envelope tests are also displayed. Here $r$-values less than $\SI{1}{\micro \meter}$ are disregarded.}
	\label{fig:JFGCoxCronie}
\end{figure}

This Appendix contains figures related to the analysis of the six spine data sets.  Figure~\ref{fig:JFGPoisson} shows $95\%$ global rank envelopes under the fitted inhomogeneous Poisson model using a concatenation of $\hat{F}$, $\hat{G}$, and $\hat{J}$ as test function, where distances less than $\SI{1}{\micro \meter}$ are disregarded. Figure~\ref{fig:RF_simulation} show one simulation of the fitted random field $\Pi$ for each network, and each of Figures~\ref{fig:neuron1_simulations}--\ref{fig:neuron6_simulations} display the data along with five simulated point patterns from the fitted Cox process model. Finally, Figure~\ref{fig:JFGCoxCronie} shows $95\%$ global rank envelopes under the fitted Cox model using a concatenation of the estimated $F$-, $G$-, and $J$-functions proposed by \cite{cronie-etal-19} as test function, where distances less than $\SI{1}{\micro \meter}$ are disregarded.

\section*{Acknowledgements}
We would like to thank Abdel-Rahman Al-Absi who collected the spine
data. We have been supported by The Danish Council for Independent
Research | Natural Sciences, grant DFF -- 7014-00074 ``Statistics for
point processes in space and beyond'', and by the ``Centre for
Stochastic Geometry and Advanced Bioimaging'', funded by grant 8721
from the Villum Foundation.

\bibliographystyle{apalike}
\bibliography{references}

\begin{thebibliography}{}

\bibitem[Anderes et~al., 2020]{anderes-etal-17}
Anderes, E., M{\o}ller, J., and Rasmussen, J.~G. (2020).
\newblock Isotropic covariance functions on graphs and their edges.

\bibitem[Ang et~al., 2012]{ang-etal-12}
Ang, Q.~W., Baddeley, A., and Nair, G. (2012).
\newblock Geometrically corrected second order analysis of events on a linear
  network, with applications to ecology and criminology.
\newblock {\em Scandinavian Journal of Statistics}, 39:591--617.

\bibitem[Baddeley et~al., 2014]{baddeley-etal-14}
Baddeley, A., Jammalamadaka, A., and Nair, G. (2014).
\newblock Multitype point process analysis of spines on the dendrite network of
  a neuron.
\newblock {\em Journal of the Royal Statistical Society: Series C (Applied
  Statistics)}, 63:673--694.

\bibitem[Baddeley et~al., 2000]{baddeley-etal-00}
Baddeley, A., M{\o}ller, J., and Waagepetersen, R.~P. (2000).
\newblock Non- and semi-parametric estimation of interaction in inhomogeneous
  point patterns.
\newblock {\em Statistica Neerlandica}, 54:329--350.

\bibitem[Baddeley et~al., 2017]{baddeley-etal-17}
Baddeley, A., Nair, G., Rakshit, S., and McSwiggan, G. (2017).
\newblock ``{S}tationary'' point processes are uncommon on linear networks.
\newblock {\em Stat}, 6:68--78.

\bibitem[Baddeley et~al., 2015]{baddeley-etal-15}
Baddeley, A., Rubak, E., and Turner, R. (2015).
\newblock {\em Spatial Point Patterns: Methodology and Applications with R}.
\newblock Chapman \& Hall/CRC Press, Boca Raton.

\bibitem[Cronie et~al., 2019]{cronie-etal-19}
Cronie, O., Moradi, M., and Mateu, J. (2019).
\newblock Inhomogeneous higher-order summary statistics for linear network
  point processes.
\newblock Available at arXiv:1910.03304.

\bibitem[Diggle, 2014]{diggle-14}
Diggle, P.~J. (2014).
\newblock {\em Statistical Analysis of Spatial and Spatio-Temporal Point
  Patterns}.
\newblock CRC Press, Lancaster, 3rd edition.

\bibitem[Guan, 2009]{guan-09}
Guan, Y. (2009).
\newblock A minimum contrast estimation procedure for estimating the
  second-order parameters of inhomogeneous spatial point processes.
\newblock {\em Statistics and Its Interface}, 2:91--99.

\bibitem[Jammalamadaka et~al., 2013]{jammalamadaka-etal-13}
Jammalamadaka, A., Banerjee, S., Manjunath, B., and Kosik, K. (2013).
\newblock Statistical analysis of dendritic spine distributions in rat
  hippocampal cultures.
\newblock {\em BMC Bioinformatics}, 14:287.

\bibitem[Lavancier and M{\o}ller, 2016]{lavancier-moeller-16}
Lavancier, F. and M{\o}ller, J. (2016).
\newblock Modelling aggregation on the large scale and regularity on the small
  scale in spatial point pattern datasets.
\newblock {\em Scandinavian Journal of Statistics}, 43:587--609.

\bibitem[Lavancier et~al., 2018]{lavancier-etal-18}
Lavancier, F., Poinas, A., and Waagepetersen, R.~P. (2018).
\newblock Adaptive estimating function inference for non-stationary
  determinantal point processes.
\newblock Available on arXiv:1806.06231.

\bibitem[Mat{\'e}rn, 1960]{matern:1960}
Mat{\'e}rn, B. (1960).
\newblock Spatial variation: Stochastic models and their application to some
  problems in forest surveys and other sampling investigations.
\newblock {\em Meddelanden fr{\aa}n Statens Skogforskningsinstitut}, 49:1--144.

\bibitem[Mat{\'e}rn, 1986]{matern:1986}
Mat{\'e}rn, B. (1986).
\newblock {\em Spatial Variation}.
\newblock Lecture Notes in Statistics 36. Springer-Verlag, Berlin.

\bibitem[McSwiggan et~al., 2016]{mcswiggan-etal-16}
McSwiggan, G., Baddeley, A., and Nair, G. (2016).
\newblock Kernel density estimation on a linear network.
\newblock {\em Scandinavian Journal of Statistics}, 44:324--345.

\bibitem[M{\o}ller et~al., 1998]{moller-98}
M{\o}ller, J., Syversveen, A., and Waagepetersen, R.~P. (1998).
\newblock Log {G}aussian {C}ox processes.
\newblock {\em Scandinavian Journal of Statistics}, 25:451--482.

\bibitem[Myllym{\"a}ki et~al., 2017]{myllymaki-17}
Myllym{\"a}ki, M., Mrkvi\v{c}ka, T., Grabarnik, P., Seijo, H., and Hahn, U.
  (2017).
\newblock Global envelope tests for spatial processes.
\newblock {\em Journal of Royal Statistical Society Series B (Statistical
  Methodology)}, 79:381--404.

\bibitem[Okabe and Sugihara, 2012]{okabe:sugihara:2012}
Okabe, A. and Sugihara, K. (2012).
\newblock {\em Spatial Analysis along Networks: Statistical and Computational
  Methods}.
\newblock Lecture Notes in Statistics 36. Wiley, Chichester.

\bibitem[Okabe and Yamada, 2001]{okabe-yamada-01}
Okabe, A. and Yamada, I. (2001).
\newblock The {$K$}-function method on a network and its computational
  implementation.
\newblock {\em Geographical analysis}, 33:271--290.

\bibitem[Rakshit et~al., 2019]{rakshit-etal-19}
Rakshit, S., Davies, T., Moradi, M., McSwiggan, G., nair, G., Mateu, J., and
  Baddeley, A. (2019).
\newblock Fast kernel smoothing of point patterns on a large network using
  two-dimensional convolution.
\newblock {\em International Statistical Review}, 87:531–556.

\bibitem[Rakshit et~al., 2017]{rakshit-etal-17}
Rakshit, S., Nair, G., and Baddeley, A. (2017).
\newblock Second-order analysis of point patterns on a network using any
  distance metric.
\newblock {\em Spatial Statistics}, 22:129--154.

\bibitem[Rasmussen and Christensen, 2019]{rasmussen-christensen-18}
Rasmussen, J.~G. and Christensen, H.~S. (2019).
\newblock Point processes on directed linear networks.
\newblock Available at arXiv:1812.09071.

\bibitem[van Lieshout, 2011]{lieshout-11}
van Lieshout, M. N.~M. (2011).
\newblock A {$J$-function} for inhomogeneous point processes.
\newblock {\em Statistica Neerlandica}, 65:183--201.

\bibitem[Waagepetersen, 2007]{waagepetersen-07}
Waagepetersen, R.~P. (2007).
\newblock An estimating function approach to inference for inhomogeneous
  {Neyman}-{Scott} processes.
\newblock {\em Biometrics}, 63:252--258.

\bibitem[Waagepetersen and Guan, 2009]{waagepetersen-guan-09}
Waagepetersen, R.~P. and Guan, Y. (2009).
\newblock Two-step estimation for inhomogeneous spatial point processes.
\newblock {\em Journal of the Royal Statistical Society Series B (Statistical
  Methodology)}, 71:685--702.

\end{thebibliography}

\end{document}